\documentclass[twocolumn, times]{aastex63}

\newcommand{\nnhp}{N$_2$H$^+$ (4-3)}
\newcommand\dcop{DCO$^+$ (5-4)}
\newcommand{\hhdp}{o-H$_2$D$^+$ (1(1,0)-1(1,1))}
\newcommand{\CS}{$^{13}$CS (8-7)}
\newcommand{\soo}{SO$_2$ 19(4,16)-19(3,17)}
\newcommand{\textsg}[1]{\textbf{#1}}

\usepackage[version=4]{mhchem}
\usepackage{multirow}
\usepackage{graphicx}
\usepackage{color}

\begin{document}

\title{\large{Chemistry in the GG Tau A Disk: Constraints from H$_2$D$^+$, N$_2$H$^+$, and DCO$^+$ High Angular Resolution ALMA Observations}}

\correspondingauthor{Liton Majumdar}
\email{liton@niser.ac.in, dr.liton.majumdar@gmail.com}

\author[0009-0008-2350-5876]{Parashmoni Kashyap}
\author [0000-0001-7031-8039] {Liton Majumdar}
\affiliation{Exoplanets and Planetary Formation Group, School of Earth and Planetary Sciences, National Institute of Science Education and Research, Jatni 752050, Odisha, India}
\affiliation{Homi Bhabha National Institute, Training School Complex, Anushaktinagar, Mumbai 400094, India}

\author[0000-0003-2341-5922]{Anne Dutrey}
\author[0000-0003-3773-1870]{St{\'e}phane Guilloteau}
\affiliation{Laboratoire d'Astrophysique de Bordeaux, Universit\'e de Bordeaux, CNRS, B18N, All\'ee Geoffroy Saint-Hilaire, F-33615 Pessac, France}

\author[0000-0001-6124-5974]{Karen Willacy} 
\affiliation{Jet Propulsion Laboratory, California Institute of Technology, 4800 Oak Grove Dr. Pasadena, CA, 91109, USA}

\author[0009-0003-6480-8084]{Edwige Chapillon}
\affiliation{Institut de Radioastronomie Millimétrique (IRAM), 300 rue de la Piscine, F-38406 Saint-Martin d’Héres, France}
\affiliation{Laboratoire d'Astrophysique de Bordeaux, Universit\'e de Bordeaux, CNRS, B18N, All\'ee Geoffroy Saint-Hilaire, F-33615 Pessac, France}

\author[0000-0003-1534-5186]{Richard Teague}
\affiliation{ Department of Earth, Atmospheric, and Planetary Sciences, Massachusetts Institute of Technology, Cambridge, MA 02139, USA}

\author[0000-0002-3913-7114]{Dmitry Semenov}
\affiliation{Max Planck Institute for Astronomy, Königstuhl 17, D-69117 Heidelberg, Germany} 
\affiliation{Department of Chemistry, Ludwig Maximilian University, Butenandtstr. 5–13, D-81377 Munich, Germany}

\author[0000-0002-1493-300X]{Thomas Henning}
\affiliation{Max Planck Institute for Astronomy, Königstuhl 17, D-69117 Heidelberg, Germany}

\author[0000-0001-8292-1943]{Neal Turner}
\author[0000-0002-6858-5063]{Raghvendra Sahai}
\affiliation{Jet Propulsion Laboratory, California Institute of Technology, 4800 Oak Grove Dr. Pasadena, CA, 91109, USA} 

\author[0000-0001-7157-6275]{\'Agnes K\'osp\'al}
\affiliation{Konkoly Observatory, HUN-REN Research Centre for Astronomy and Earth Sciences, CSFK, MTA Centre of Excellence, Konkoly-Thege Mikl\'os \'ut 15-17, 1121 Budapest, Hungary}
\affiliation{Institute of Physics and Astronomy, ELTE E\"otv\"os Lor\'and University, P\'azm\'any P\'eter s\'et\'any 1/A, 1117 Budapest, Hungary}
\affiliation{Max Planck Institute for Astronomy, Königstuhl 17, D-69117 Heidelberg, Germany}

\author[0000-0003-1805-3920]{Audrey Coutens}
\affiliation{Institut de Recherche en Astrophysique et Planétologie (IRAP), Université de Toulouse, UT3-PS, CNRS, CNES, 9 av. du Colonel Roche, 31028 Toulouse Cedex 4, France}
\author[0009-0006-3497-397X]{V. Piétu}
\affiliation{Institut de Radioastronomie Millimétrique (IRAM), 300 rue de la Piscine, F-38406 Saint-Martin d’Héres, France}

\author[0000-0002-6636-4304]{Pierre Gratier}
\affiliation{Laboratoire d'Astrophysique de Bordeaux, Universit\'e de Bordeaux, CNRS, B18N, All\'ee Geoffroy Saint-Hilaire, F-33615 Pessac, France}

\author[0000-0003-0522-5789]{Maxime Ruaud}
\affiliation{NASA Ames Research Center, Moffett Field, CA 94035, USA}
\affiliation{Carl Sagan Center, SETI Institute, Mountain View, CA 94035, USA}

\author[0000-0002-4372-5509]{N. T. Phuong}
\affiliation{Department of Astrophysics, Vietnam National Space Center, Vietnam Academy of Science and Techonology, 18 Hoang Quoc Viet, Cau Giay, Hanoi, Vietnam}

\author[0009-0009-9618-4927]{E. Di Folco}
\affiliation{Laboratoire d'Astrophysique de Bordeaux, Universit\'e de Bordeaux, CNRS, B18N, All\'ee Geoffroy Saint-Hilaire, F-33615 Pessac, France} 

\author[0000-0002-3024-5864]{Chin-Fei Lee}
\author[0000-0002-0675-276X]{Y.-W. Tang}
\affiliation{Academia Sinica Institute of Astronomy and Astrophysics, PO Box 23-141, Taipei 106, Taiwan}

\begin{abstract}
    Resolved molecular line observations are essential for gaining insight into the physical and chemical structure of protoplanetary disks, particularly in cold, dense regions where planets form and acquire their chemical compositions. However, tracing these regions is challenging because most molecules freeze onto grain surfaces and are not observable in the gas phase. We investigated cold molecular chemistry in the triple stellar T Tauri disk GG Tau A, which harbours a massive gas and dust ring and an outer disk, using ALMA Band 7 observations. We present high angular resolution maps of \ce{N2H+} and \ce{DCO+} emission, with upper limits reported for \ce{H2D+}, $^{13}$CS, and \ce{SO2}. The radial intensity profile of \ce{N2H+} shows most emission near the ring outer edge, while \ce{DCO+} exhibits double peaks, one near the ring inner edge and the other in the outer disk. With complementary observations of lower-lying transitions, we constrained the molecular surface densities and rotation temperatures. We compared the derived quantities with model predictions across different cosmic ray ionization (CRI) rates, carbon-to-oxygen (C/O) ratios, and stellar UV fluxes. Cold molecular chemistry, affecting \ce{N2H+}, \ce{DCO+}, and \ce{H2D+} abundances, is most sensitive to CRI rates, while stellar UV flux and C/O ratios have minimal impact on these three ions. Our best model requires a low cosmic ray ionization rate of $10^{-18}$ s$^{-1}$. However, it fails to match the low temperatures derived from \ce{N2H+} and \ce{DCO+}, 12 to 16 K, which are much lower than the CO freezing temperature.
\end{abstract}


\section{Introduction} \label{sec:intro}
A protoplanetary disk is a crucial stage in the evolution of cosmic matter, where the interplay of inherited material from a parent cloud and the in-situ physics and chemistry establishes the conditions for planet formation. Protoplanetary disks are characterized by complex chemical stratification owing to radial and vertical variations in density, temperature, ionization, and dissociative radiation \citep{Aikawa2002}. The uppermost layers usually allow the formation of simple atoms, ions, and photo-stable radicals, while the intermediate warm molecular layer produces most of the gas phase molecules traced by the  Atacama Large Millimeter/submillimeter Array (ALMA) observations. There is minimal penetration of stellar radiation in the innermost regions near the midplane in the vertical direction, making them very cold and poorly ionized. In this dense region, most of the molecules freeze onto grain surfaces. Hence, midplane regions are very difficult to probe. N$_2$H$^+$, DCO$^+$, and H$_2$D$^+$ are among the very few molecules which remain in the gas phase in the midplane region \citep{Ceccarelli2004, Oberg2011}.\par

N$_2$H$^+$ is expected to be abundant in regions where \ce{CO} is depleted. This anti-correlation is observed in both dense cores \citep[e.g.,][]{Caselli1999, Bergin2002} as well as in disks \citep{Walsh2012}. \ce{N2H+} is produced through the protonation of \ce{N2} and rapidly destroyed by CO in its presence to form \ce{HCO+}. 
Hence, N$_2$H$^+$ can persist in the gas phase at temperatures a few Kelvin below the CO freeze-out temperature \citep{Oberg2005}. N$_2$H$^+$ has been observed in multiple disks \citep[e.g.,][]{Dutrey2007, Qi2013, Phuong2021, Anderson2022}, but is often characterized by comparatively weak emission.\par

An enhanced abundance of H$_2$D$^+$ is predicted by chemical models at low temperatures (T$<$20 K), with the major formation pathway being protonation of \ce{HD} \citep{Robert2000}. H$_2$D$^+$ can only be observed in high-density regions with low \ce{CO} and \ce{N2} content where \ce{H3+} has not already been destroyed. H$_2$D$^+$ has already been detected in prestellar cores \citep[e.g.,][]{Caselli2008}, but there is no clear detection in any of the protoplanetary disks \citep[e.g.,][]{Ceccarelli2004,Thi2004,Qi2008, Chapillon2011}. \par

DCO$^+$ serves as an effective indicator of low-temperature CO freezeout regions within the disk and is predominantly formed through the interaction of \ce{CO} with H$_2$D$^+$ \citep{Wootten1987}. On the one hand, the formation of \ce{H2D+} mainly occurs through HD protonation and is hindered by the destruction of \ce{H3+} in the presence of CO. On the other hand, the presence of CO is essential for DCO$^+$ formation. Consequently, DCO$^+$ emission is expected to peak in a narrow region around the \ce{CO} snowline where both the parent molecules are in the gas phase in small amounts \citep{Aikawa2002, Willacy2007, Mathews2013}.\par

In summary, N$_2$H$^+$, DCO$^+$, and H$_2$D$^+$ are excellent tracers of low-temperature, high-density regions of the disk, specifically the midplane region in outer disks. Simultaneous observations of these molecules allow us to locate the CO snowline. Snowlines are thought to play an important role in planet formation because the ice coating on grains increases the solid mass, potentially promotes the coagulation of grains into larger particles, and/or the larger bodies formed from these grains require a higher collision velocity for destruction. These effects are particularly prominent just outside the snowline \citep{Okuzumi2012}. Additionally, such observations provide insights into ionization fractions within the relevant disk areas, offering constraints that can impact magneto-rotational instability. In instances of excessively low ionization fractions, ``dead zones" \citep{Gammie1996} form, which are conducive to efficient grain growth and subsequent planet formation \citep[e.g.,][]{Oberg2011}. \par

This work reports spatially resolved emission from \nnhp\ and \dcop\ and upper limits for \hhdp, \CS, and \soo\ in the disk around a triple stellar T Tauri system, GG Tau A, utilizing the ALMA interferometer. This represents the most sensitive observation of these molecules in GG Tau A to date. The system (age 1-5 Myr) is located at a distance of 150 pc \citep[using the GAIA parallax measurements of the GG Tau Ba star of this hierarchical quintuple system, because those of GG Tau A are contaminated by its triple nature,][]{Gaia2016,Gaia2018} from us in the Taurus-Auriga star-forming region. This circumtertiary disk is characterized by a dense ring of gas and dust between 193 and 285 au (from here on, we simply use the term ring to denote this region) where 90\% of the continuum emission arises and an outer disk extended up to 800 au in $^{13}$CO observations \citep{Guilloteau1999, Dutrey2014}. The majority of the total disk mass (0.15 $M_{\odot}$) is concentrated in the ring region (0.13 $M_{\odot}$) \citep{Guilloteau1999, Andrews2014}. \citet{Dutrey2014} reported the presence of a hot spot located at the outer edge of the dense ring where a partial CO gap is observed \citep{Tang2016}, suggesting the presence of an embedded planet. This hypothesis is reinforced by the detection of a CO spiral emanating from the hot spot \citep{Phuong2020a}.

This disk is very cold, with a midplane temperature of $\sim$ 14 K \citep{Dutrey2014} and an atmospheric temperature of $\sim$ 30 K \citep{Guilloteau1999, Phuong2020}, at a radial distance of 214 au from the center. The first-ever detection of \ce{H2S} in the outer disk of GG Tau A by \citet{Phuong2018} further confirms how dense and massive this system is. Given its substantial mass, high density (which should ease the detection of weak molecular lines), and low temperature, GG Tau A emerges as an ideal candidate for unravelling the intricacies of cold molecular chemistry within a protoplanetary disk. \par

We present our observational details in Section \ref{sec:obs} of this paper. We explain the schemes we undertake to reduce the data in Section \ref{subsec: data_red} and the observational results in Section \ref{subsec:obs_res}, including integrated intensity maps, radial profiles, and Keplerian deprojected spectra. We also constrain the surface density of the detected molecules and the upper limits of surface density for the undetected molecules, as explained in Section \ref{sec:constr_disk_phy_params}. We perform a chemical analysis to understand our observations better using our astrochemical model, detailed in Section \ref{sec:models}. We discuss our results in Section \ref{sec:discussions} and summarize the key understandings of our study in Section \ref{sec:conclusion}.

\section{Observations}
\label{sec:obs}

\begin{deluxetable*}{ccccccc}[!ht]
\tablecaption{Spectral Setup of the ALMA Band 7 Observations of GG Tau A}
\label{tab:spectral_setup}
\tablehead{
  \colhead{\multirow{2}{*}{Spectral Window}} &
  \colhead{\multirow{2}{*}{Central Frequency (GHz)}} &
  \multicolumn{2}{c}{Bandwidth} &
  \multicolumn{2}{c}{Resolution} &
  \colhead{\multirow{2}{*}{No. of Channels}} \\
  & & \colhead{Frequency (MHz)} & \colhead{Velocity (km/s)} & \colhead{Frequency (MHz)} & \colhead{Velocity (km/s)} & 
}
\startdata
25 & 358.000000 & 1875.00 & 1570.2 & 1.129 & 0.945 & 3840 \\
27 & 359.770685 & 117.19 & 97.7 & 0.141 & 0.118 & 1920 \\
29 & 360.169780 & 117.19 & 97.5 & 0.141 & 0.117 & 1920 \\
31 & 369.908554 & 117.19 & 95.0 & 0.141 & 0.114 & 1920 \\
33 & 372.421385 & 117.19 & 94.3 & 0.122 & 0.098 & 1920 \\
35 & 372.672493 & 58.59 & 47.1 & 0.141 & 0.114 & 960
\enddata
\end{deluxetable*}

The GG Tau A system (ICRS 04:32:30.3460, 17:31:40.642) was observed as part of the ALMA project \#2021.1.00342.S (PI: Liton Majumdar) targeting to investigate the presence of Band 7 transitions, \hhdp, \nnhp, \dcop\, \CS\ and \soo. The observations were made over two executions using 47 antennas with baselines ranging from 14 m to 456 m for a total on-source integration time of 2.7 hours. The spectral setup details can be found in Table \ref{tab:spectral_setup}.

\subsection{Data Reduction} \label{subsec: data_red}
Calibration procedures involved bandpass and flux calibration, with the quasar source J0423-0120 utilized for this purpose. Phase calibration was conducted using the sources J0510+1800 and J0440+1437. The observed data were initially calibrated with CASA, the Common Astronomy Software Applications package (version 6.4.1.12) pipeline. Subsequently, five rounds of continuum self-phase calibration were performed for each execution, selecting line-free channels carefully. The self-calibration solutions were then applied to the entire dataset. To determine the disk center, we fitted the continuum structure with an ellipse, and both executions were recentered to the disk center using the \texttt{fixvis} task in CASA. The offset between the disk centers in the two executions was only (0.005$''$, 0.003$''$), which is well below our observed angular resolution of 0.5$''$– 0.9$''$. The data with the identical spectral setups from both executions were then combined using the \texttt{concat} task in CASA. We performed continuum subtraction on the self-calibrated recentered data by fitting a first-order polynomial to the continuum channels using \texttt{uvcontsub} task in CASA. We constructed the images from our visibility dataset in  \texttt{tclean} task in CASA using \texttt{hogbom} deconvolving algorithm with \texttt{briggs} weighting (robust=1.0). This yielded an angular resolution of $0.93'' \times 0.77''$ at Position Angle (PA) $68.76^\circ$, and a typical sensitivity of 14.6 mJy/beam for N$_2$H$^+$ (channel maps are presented in Figure \ref{fig:ch_maps_n2hp}). For DCO$^+$, we achieved an angular resolution of $0.91'' \times 0.73''$
at PA $99.20^\circ$, with a sensitivity of 5.6 mJy/beam (channel maps are presented in Figure \ref{fig:ch_maps_dcop}). For H$_2$D$^+$, the noise (16.7 mJy/beam) is higher as the atmospheric transmission is degraded at this high frequency.
For the weaker lines, including H$_2$D$^+$, we have used \texttt{multiscale} deconvolver with \texttt{scales = [0,5,15,25]} $\times$ pixels where each pixel corresponds to one-seventh of the beam size. We adopted the \texttt{natural} weighting scheme to enhance the signal-to-noise ratio for the weak lines. The masks used during cleaning were generated with the \texttt{keplerian\_mask.py} tool \footnote{\url{https://github.com/richteague/keplerian\_mask}} with the geometric parameters $M_{*}$ = $ 1.37 M_{\odot}$, i = 37$^{\circ}$, and PA = 278$^{\circ}$ confined within inner and outer radii of 1$''$ and 3.5$''$ respectively (values rescaled from \citet{Guilloteau1999} for a distance of 150 pc). Additionally, a continuum image was generated for a spectral window with a central frequency of 358 GHz, utilizing the line-free channels. The deconvolution algorithm employed was \texttt{hogbom}, and the process utilized the \texttt{briggs} weighting scheme with the robust parameter set to 0.

\subsection{Observational Results} \label{subsec:obs_res}
\begin{figure*}[ht!]
    \centering
    \includegraphics[width=\textwidth]{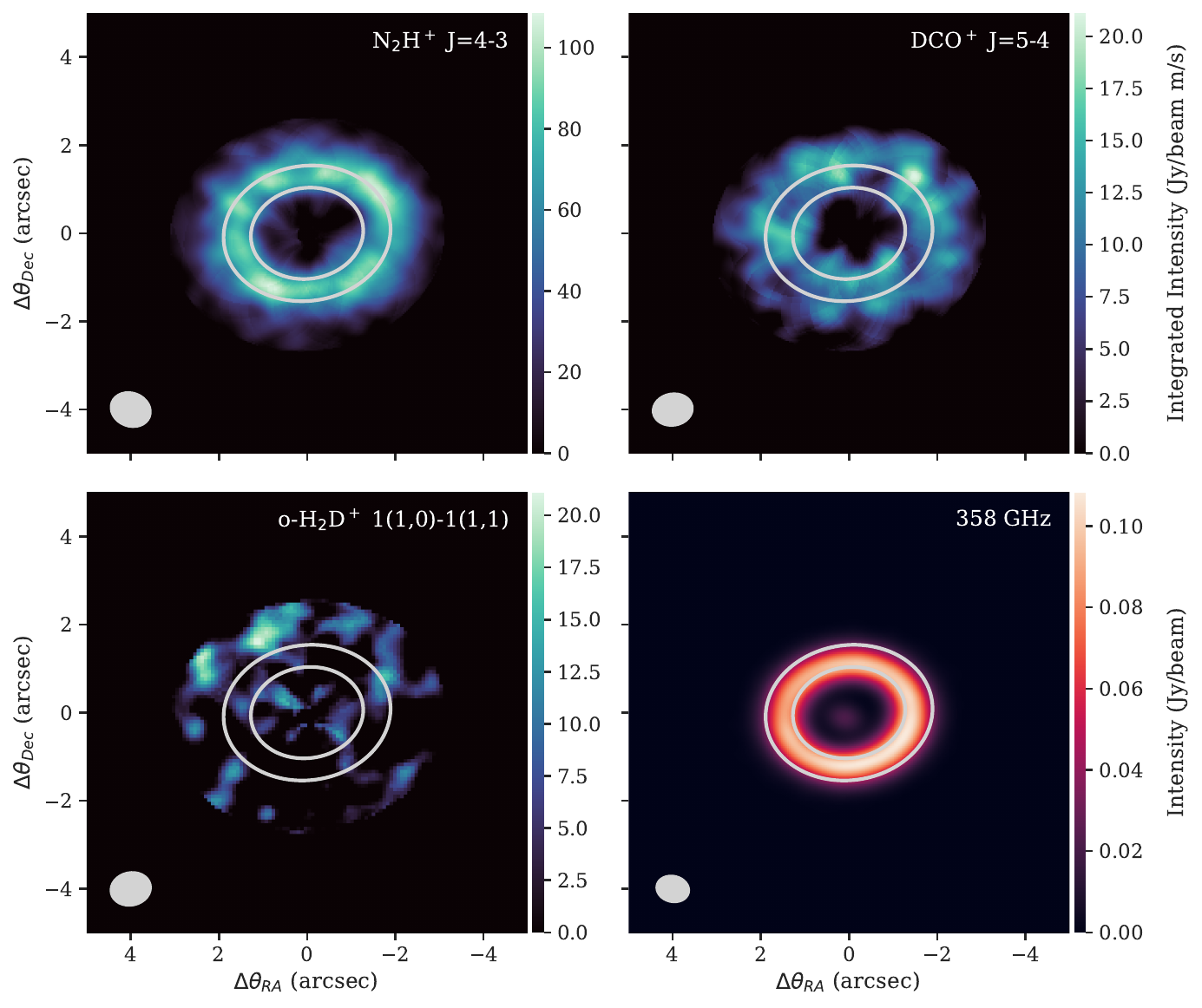}
    \caption{Integrated intensity maps of the targeted lines \nnhp, \dcop\ and \hhdp. The overplotted elliptical ring contours in white are at 193 au (1.28$''$) and 285 au (1.9$''$) to indicate the continuum ring region. Both N$_2$H$^+$ and DCO$^+$ emissions are extended beyond the ring region. As far as H$_2$D$^+$ is concerned, there is no significant emission from the ring region. The figure at the lower right corner is the continuum image at 358 GHz.
    }
    \label{fig:mo_maps}
\end{figure*}

The integrated intensity maps were generated by collapsing the image cubes along the velocity axis while applying the Keplerian masks used for cleaning. We did not use any intensity threshold criteria to decide the inclusion of pixels while generating the moment map. For this purpose, we employed \texttt{bettermoments} python package \citep{2018RNAAS...2c.173T}. Figure \ref{fig:mo_maps} displays integrated intensity maps for the targeted molecules, where the elliptical contours indicate the location of the dust continuum ring region (193 au - 285 au) of the disk GG Tau. It also displays the continuum image at the central frequency of 358 GHz in the bottom right corner. Clear detections of \nnhp\ and \dcop\ are observed, while \hhdp\, \CS\ and \soo\ show no detection.\par

\begin{figure*}[!ht]
    \centering
    \includegraphics[width=\textwidth]{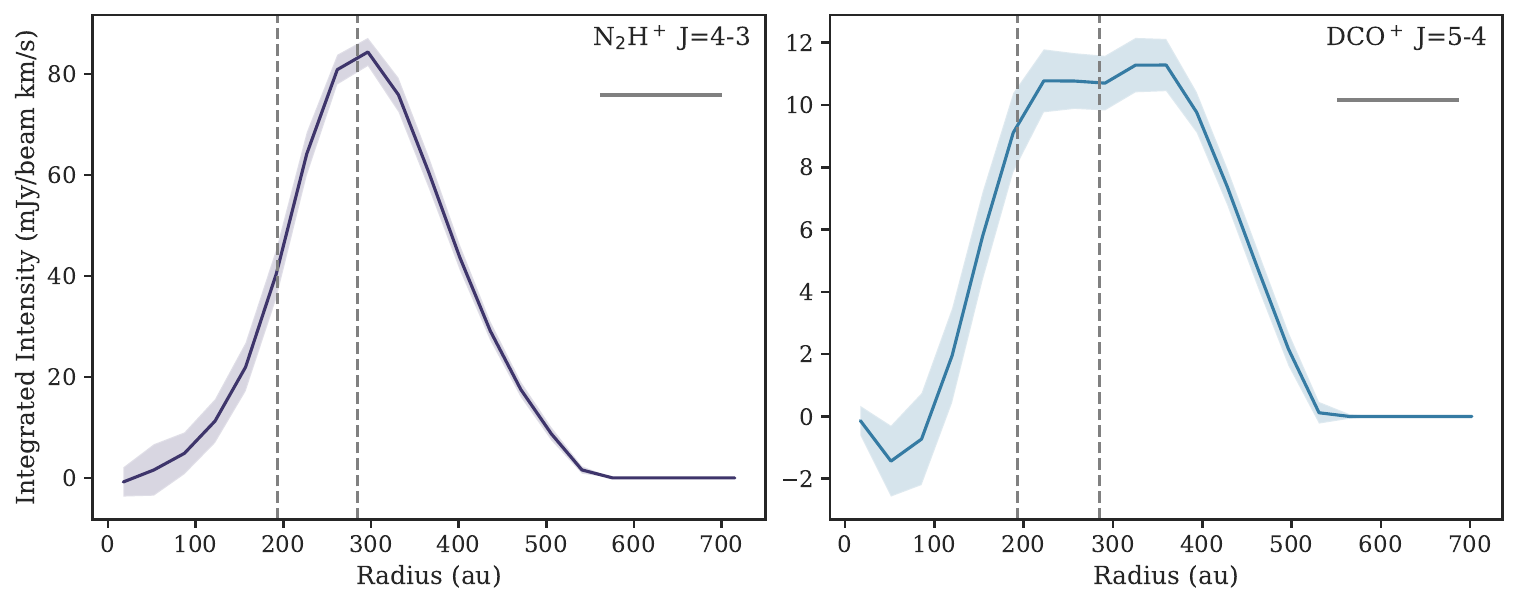}
    \caption{Azimuthally averaged velocity integrated radial profiles of the detected lines generated from the integrated intensity maps shown in Figure \ref{fig:mo_maps}. The shaded region in the above figure corresponds to 1$\sigma$. The dashed vertical lines are to designate the location of the ring around GG Tau A, spanning from 193 au to 285 au. The grey horizontal line at the top right corner represents the beam major axis.}
    \label{fig:radial_intensity}
\end{figure*}

We generated radial distribution of velocity integrated brightness across the disk from the zeroth moment maps with the help of \texttt{radial\_profile} function in \texttt{GoFish} python package \citep{GoFish} by dividing the disk radial range into a series of annular rings with a width 1/4th of the beam major axis and averaging the emission for each ring. The radial profiles for \nnhp\ and \dcop\ are shown in Figure \ref{fig:radial_intensity}. The two vertical lines indicate the ring region (193 au - 285 au). We have also shown the radial profiles for all the transitions generated without keplerian masking in Figure \ref{fig:rad_profile_wihtout_kep}. \ce{DCO+} shows a slightly lower emission in the ring region in this case.\par

\begin{figure*}[t!]
    \centering
    \includegraphics[width=\textwidth]{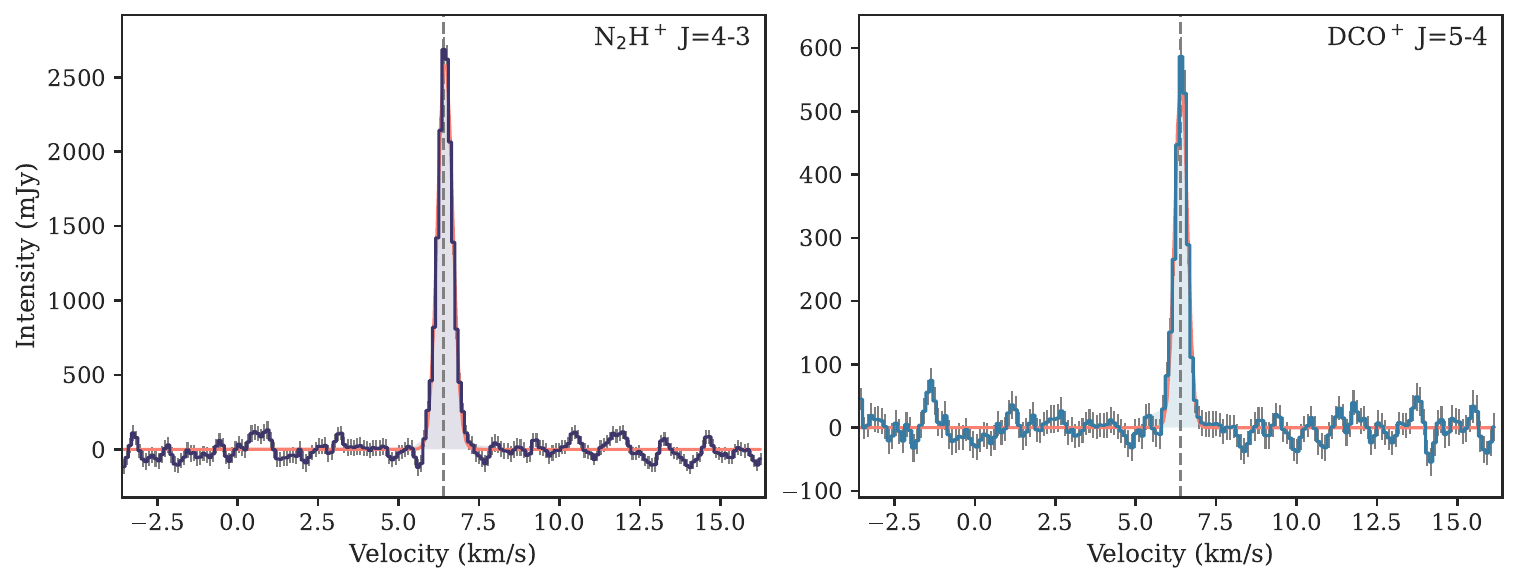}
    \caption{Keplerian deprojected and stacked spectra of the detected lines \nnhp\ and \dcop. The shaded region is the region over which we have done a Gaussian fitting (orange curve) and calculated integrated flux densities. The stacked spectra are peaking at the systemic velocity, $v_{\text{LSRK}}$ of 6.4 km/s, confirming the detections. The location of the systemic velocity is shown by the vertical dashed lines. Integrated flux densities and matched filter responses are tabulated in Table \ref{tab:obs_lines}.}
    \label{fig:int_spectra}
\end{figure*}

We employed the shifting and stacking technique using the \texttt{integrated\_spectrum} function in the \texttt{GoFish} package \citep{GoFish} to generate a disk-integrated spectrum. Velocities for each pixel were shifted to the systemic one, 6.4 km/s, utilizing system kinematic information used in the generation of the Keplerian mask and stacked to create a Keplerian deprojected spectrum for each transition. The radial range for generating the integrated spectrum is between 120 au and 550 au. We have done a Gaussian fit to the observed spectrum using the \texttt{scipy.optimize.curve\_fit} to determine the peak intensity and the velocity at which the peak occurs. The width of the observed signal is full width at half maximum, given by $2\sqrt{2 \ln{2}}\sigma$, where $\sigma$ is the standard deviation calculated from our fit. We used \texttt{numpy.trapz} function to calculate the disk-integrated flux density of the line signal over a velocity range determined by visual inspection (the shaded region in Figure \ref{fig:int_spectra}). The Keplerian deprojected spectra are shown in the upper panel of Figure \ref{fig:int_spectra}. The stacked spectra for \nnhp\ and \dcop\ are peaking at the systemic velocity, $v_{\text{LSRK}}$ of 6.4 km/s, confirming the detections. The disk-integrated flux densities are reported in Table \ref{tab:obs_lines}. The $1\sigma$ errors are calculated as $f \times \delta S \times \delta v \times \sqrt{N}$, where $f$ is the spectral correlation factor derived from the noise autocorrelation spectrum), $\delta S$ is the noise associated with each channel, $\delta v$ is the channel width, and $N$ is the number of channels over which integration is performed.\par

\begin{deluxetable*}{lcccccccc}
\tablecaption{Properties of the Lines Observed}
\label{tab:obs_lines}
\tablewidth{\textwidth}
\tablehead{
  \colhead{Species} & \colhead{Transition} & \colhead{Frequency} & \colhead{E$_u$} & \colhead{log$_{10}$(A$_{ul}$$)$} & \colhead{g$_u$} & \colhead{Filter Response} & \colhead{Integrated Intensity} &
  \colhead{Line Width}\\[-1ex]
  \colhead{} & \colhead{} & \colhead{(GHz)} & \colhead{(K)} & \colhead{(s$^{-1}$)} & \colhead{} & \colhead{($\sigma$)}
  & \colhead{(mJy km s$^{-1}$)} & \colhead{(km s$^{-1}$)}
}
\startdata
N$_2$H$^+$ & 4-3 & 372.6724808 & 44.71 & -2.50934 & 81 & 43 & 1570 $\pm$ 40 & 0.55 $\pm$ 0.01 \\
DCO$^+$ & 5-4 & 360.1697783 & 51.86 & -2.42480 & 11 & 20 & 290 $\pm$ 20 & 0.46 $\pm$ 0.01 \\
H$_2$D$^+$ & 1(1,0)-1(1,1) & 372.4213558 & 104.20 & -3.96567 & 9 & - & \textless{}58 & - \\
$^{13}$CS & 8-7 & 369.9085505 & 79.89 & -2.97351 & 34 & - & \textless{}70 & - \\
SO$_2$ & 19(4,16)-19(3,17) & 359.7706846 & 214.26 & -3.41473 & 39 & - & \textless{}35 & - \\
\enddata
\tablecomments{
Here, E$_{u}$ is the upper energy level of the transition, A$_{ul}$ is the Einstein coefficient for spontaneous transition, and g$_u$ is the upper state degeneracy. The error bars on the integrated intensity for the detected molecules are 1$\sigma$. We have calculated a 3$\sigma$ upper limit on flux densities for the undetected molecules for an integration area between radius 150 au and 450 au and a line width of 0.55 km/s (Appendix \ref{sec:rotational_diagram}). The line width comes from the full-width half maximum of the Gaussian fitting in Figure \ref{fig:int_spectra} (Section \ref{subsec:obs_res}).}
\end{deluxetable*}
%
We further verified our detections with the matched filtering technique \citep{loomis2018detecting}. In this method, we cross-correlate observed visibility with modelled visibility derived from the Keplerian masks used during cleaning. We utilized a package called \texttt{VISIBLE} \citep{loomis2018visible} for this task. For \nnhp, we found a filter response of 43\,$\sigma$, and for \dcop, it is 20\,$\sigma$ at the systemic velocity of 6.4 km/s (Table \ref{tab:obs_lines}). The matched filter responses are shown in the Figure \ref{fig:matched_filter}.

Additionally, calibrated visibilities were exported to UVFITS format using the CASA task \texttt{exportuvfits} for subsequent analysis with the \textsc{Imager} program\footnote{\url{https://imager.oasu.u-bordeaux.fr}} and the \textsc{DiskFit} disk model fitting tool \citep{Pietu2007}. Spectra and radial peak brightness profiles generated by \texttt{KEPLER} are presented in Figure \ref{fig:kepler-n2h} in Appendix \ref{app:imager}. The \texttt{KEPLER} command in \textsc{Imager} is similar to the \texttt{GoFish} package.
The reconstructed spectra and radial profiles slightly differ from those in Figure \ref{fig:int_spectra} because of the higher angular resolution ($0.65'' \times 0.53''$ at PA $80^\circ$) used in \textsc{Imager}.

Both \nnhp\ and \dcop\ are detected with a very high signal-to-noise ratio, with \nnhp\ being the stronger detection among the two. Radial profiles (Figures \ref{fig:radial_intensity}, \ref{fig:kepler-n2h}) indicate that the \nnhp\ emission exhibits a ring-like radial distribution, peaking near the outer edge of the continuum ring. In contrast, the radial distribution of \dcop\ peak brightness temperature (Figure \ref{fig:kepler-n2h}) shows a double-ringed structure. The inner \dcop\ emission arises from the continuum ring, while an outer, brighter \dcop\ emission originates from the outer disk region. This suggests radial variations in the deuteration process across the disk. However, this double-ringed structure is less prominent in the velocity-integrated radial profile (right panel in Figure \ref{fig:radial_intensity}), possibly due to the lower angular resolution. Other targeted species, including H$_2$D$^+$, \ce{$^{13}$CS}, and SO$_2$, are not detected (Figure \ref{fig:matched_filter}).

\section{Derivation of Disk Physical Parameters} 
\label{sec:constr_disk_phy_params}

\subsection{DiskFit modelling}
\label{subsec:diskfit}
To derive disk properties, we perform a least-square fit to the observed visibilities to adjust an empirical parametric disk model to the data, using the \textsc{DiskFit} tool from \citet{Pietu2007}. This approach allows an accurate derivation of the geometrical parameters. Furthermore, when several transitions of the same molecules are available, it allows a quantitative estimate of the molecular surface densities and excitation temperatures. It also allows for consistent estimates of the errors on the disk parameters (within the framework of the adopted disk model). This can be done either from the covariance matrix (when parameter coupling is limited) or through a more elaborate
Monte Carlo Markov Chain method.

The disk model is similar to that used in \citet{Phuong2020}. It uses a flared disk geometry with a Gaussian vertical profile and a radial power-law for the scale height. Molecules are assumed to follow the same vertical distribution for simplicity. The model assumes level populations are governed by a Boltzmann law, with the temperature being a power law of radius. For a simple comparison with the observed data, we assume that the intrinsic line width remains constant with radius. In fact, fitting line width as a power law instead suggests a small but non-significant decrease of intrinsic line width with radius that does not affect the other disk parameters. The emerging line intensity is derived using simple ray tracing, considering the geometry and dynamics  (position, orientation, inclination, velocity, and Keplerian rotation pattern) and level populations computed at every sampled point along the line of sight. The hyperfine structure of the observed lines is also considered at this stage. This approach implicitly accounts for line optical depth. 
Visibilities are then computed on the same $(u,v)$ points as the data from the 3D data cube. A least square minimization is performed on the different visibilities using a modified Levenberg-Marquardt algorithm. Guided by the radial distribution of velocity-integrated brightness (Figure \ref{fig:radial_intensity}) and apparent brightness temperature (Figure \ref{fig:kepler-n2h}), we represented the radial surface density profiles for each molecule by the sum of two Gaussians, truncated to inner and outer radii of 150 and 550 au, respectively. The rotation temperature is assumed to be a simple power law, defined by its value at 250 au, $T_{250}$ and exponent $q$. The model thus has 15 possible parameters, 6 due to geometry and 9 to represent the molecular distribution and excitation conditions (see Tables \ref{tab:geom}-\ref{tab:diskfitabun} for their designation). 

Error bars are derived from the covariance matrix. The six geometric parameters, the line width $\delta V$, and temperature parameters exhibit very little coupling among them and with the other ones. However, the Gaussian distribution parameters are strongly coupled, particularly the pair ($N_i, W_i$) for distribution $i$, so their error bars should be treated with caution.\par 
The parameter fitting process involves a two-stage approach. In the first stage, geometric parameters are individually fitted for every observed transition.
The geometric parameters derived from N$_2$H$^+$ and DCO$^+$ 
are given in Table \ref{tab:geom}, using the distance
of 150 pc determined from Gaia. These transitions are relatively optically thin and, thus, more reliable tracers of the disk inclination than those previously used. Derived inclinations and orientation agree (but offer better precision) with previous determinations from CO isotopologues  \citep{Dutrey2014,Phuong2018}.
\par
The second stage entails averaging the corresponding values of six geometric parameters across both spectral lines, fixing these averaged values, and proceeding to fit the remaining nine parameters.

\begin{deluxetable}{lrclrclc}
\tablecaption{Geometric and dynamic parameters determined through \textsc{DiskFit} modelling}
\tablewidth{\columnwidth}
\tablehead{
  \colhead{Quantity} & \colhead{N$_{2}$H$^{+}$ (4-3)} &  \colhead{DCO$^{+}$ (5-4)} &  Adopted 
  }
\startdata
X$_0$ ($''$) &   -0.007 $\pm$     0.011 &    0.019  $\pm$  0.016 &  0 \\
Y$_0$ ($''$) &   -0.001 $\pm$     0.007 &    0.004  $\pm$  0.011 & 0 \\
PA ($^\circ$) &      8.4 $\pm$   0.5 &  8.1  $\pm$   0.5 & 8 \\
$i$ ($^\circ$) &     36.5 $\pm$  0.4 &  37.0  $\pm$  0.4 & 37 \\
$V_{sys}$ (km.s$^{-1}$) &  6.45 $\pm$  0.03 &     6.43 $\pm$  0.03 & 6.43 \\
$V_{100}$  (km.s$^{-1}$) & 3.61 $\pm$  0.03 &     3.54 $\pm$  0.03 & 3.55 
\enddata 
\label{tab:geom}
\tablecomments{
\newline{}
Best fit geometric parameters from the observed visibilities. Offsets (X$_0$,Y$_0$) are from the ring center. $V_{100}$ is the Keplerian velocity at 100 au. PA of 8$^\circ$ in \textsc{DiskFit} is equivalent to 278$^\circ$ in \texttt{GoFish}. In \textsc{DiskFit}, PA is measured with respect to the disk minor axis, while in \texttt{GoFish}, this angle is measured with respect to the disk major axis. The ``Adopted'' column
indicates the fixed parameter values used for the multi-transition model fit in Table \ref{tab:diskfitabun}}

\end{deluxetable}

\begin{figure*}[!ht]
    \centering
    \includegraphics[width=\textwidth]{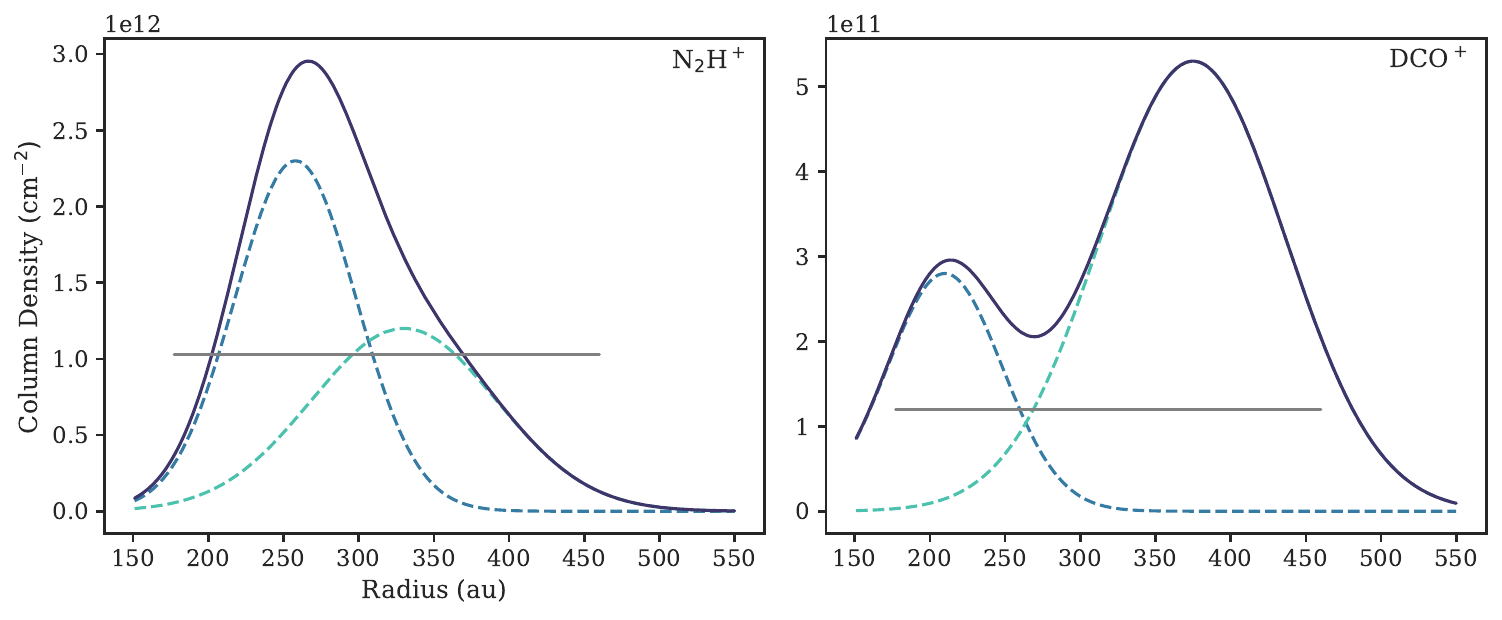}
\caption{Radial distribution of surface density derived from the \textsc{DiskFit} modelling. The solid purple lines represent the total surface density while the dashed lines show the contribution from the individual Gaussians fitted in \textsc{DiskFit}. The grey horizontal lines indicate the derived disk-averaged values using the simple LTE approach.}
\label{fig:diskfit}
\end{figure*}

To derive the surface densities and rotation temperatures, we used complementary data, which is made of the interferometric visibilities obtained with NOEMA for N$_2$H$^+$ (1-0) and DCO$^+$ (1-0) from  \cite{Phuong2021}, and the DCO$^+$ (3-2) from \cite{Phuong2018}. Geometric parameters from the ``Adopted" column of Table \ref{tab:geom} were used at this stage. Results are presented in Table \ref{tab:diskfitabun}. Figure \ref{fig:diskfit} shows the constrained radial distributions of surface densities. Agreement with the observations can be seen in Figure \ref{fig:kepler-n2h}, where the best fit radial profile (reconstructed by imaging the model visibilities and applying the Kepler deprojection) is compared with the data. \par
While this process is appropriate for DCO$^+$, 
the N$_2$H$^+$ distribution can almost equally well be represented by a single Gaussian or a truncated power law. This, however, does not affect the temperature derivation. The temperature that best represents N$_2$H$^+$ is almost constant, slightly increasing with radius and quite low, 12 K. On the contrary, DCO$^+$ is better represented by higher temperatures, decreasing with radius from $\sim 18$\,K at 200 au to $\sim 12$ K at 500 au.\par

As expected, the FWHM line width derived from the $uv$ plane analysis with \textsc{DiskFit} (0.25 \,km/s using the 1/e values in Table\,\ref{tab:diskfitabun}) is smaller than those in the recentered spectra from \texttt{GoFish} (0.55 km/s, Fig.\ref{fig:int_spectra}) and \texttt{KEPLER} (0.37 km/s, Fig. \ref{fig:kepler-n2h}) because of the remaining Keplerian shear within the synthesized beams used in these image plane reconstructions.

For H$_2$D$^+$, we simply assumed the same geometrical parameters (``Adopted'' column in Table \ref{tab:geom}), a constant temperature of 15\,K, and a uniform surface density between 150 and 450 au. The best fit surface density of ortho-H$_2$D$^+$ from \textsc{DiskFit} is then $(0.7 \pm 0.7) \times 10^{10}$ cm$^{-2}$. This gives the 3$\sigma$ upper limit on o-\ce{H2D+} to be $2.1 \times 10^{10} $ cm$^{-2}$.
%

\begin{deluxetable}{lccl}[!hbt]
\tablecaption{Best fit parameters for the
molecular distributions and excitation conditions.} \tablewidth{\columnwidth}
\tablehead{
  \colhead{Quantity} & \colhead{N$_{2}$H$^+$} & \colhead{DCO$^{+}$} &  Unit 
  }
\startdata
$\delta V$ &    0.149 $\pm$ 0.005 &    0.150 $\pm$  0.005 & km\,s$^{-1}$\\
$T_{250}$ &     12.2 $\pm$  0.7  &     16.0  $\pm$  1.2   & K\\
$q$       &    -0.32  $\pm$ 0.19 &     0.39 $\pm$  0.15   & \\
$N_{1}$   &   22 $\pm$ 8    &    2.8 $\pm$ 0.6  & $10^{11}$ cm$^{-2}$ \\
$R_{1}$   &    252 $\pm$   9 &    210 $\pm$  50  & au \\
$W_{1}$   &     95 $\pm$ 115 &     90 $\pm$ 360  & au \\
$N_{2}$   &   12 $\pm$   2   &   5.3  $\pm$ 1.0  & $10^{11}$ cm$^{-2}$\\
$R_{2}$   &    336 $\pm$ 20 &    375 $\pm$ 10  & au\\
$W_{2}$   &    145 $\pm$ 15 &     90 $\pm$ 15  & au\\
\enddata 
\label{tab:diskfitabun}
\tablecomments{
\newline
Parameters derived from the best fit of a double Gaussian radial distribution model.
$N_{i}$ is the peak surface density of Gaussian distribution $i$, $R_{i}$ its peak position, and $W_{i}$ its width (FWHM). $\delta V$ is the 1/e line width. $T_{250}$ is the temperature at 250 au, $q$ the exponent of the temperature power law \textbf(a positive exponent indicates a quantity decreasing
with radius)}.
\end{deluxetable}

\subsection{Results}
\label{sec:surfden_results}
Figure \ref{fig:diskfit} displays the radial distribution derived from the \textsc{DiskFit} modelling. The results from a classical LTE approach on disk averaged quantities (presented in  Appendix \ref{sec:rotational_diagram}) are also indicated. The discrepancy between radial profile and disk averaged value for DCO$^+$ at large radii can be attributed to the assumption of constant $T_{rot} = 16$\, K while the \textsc{DiskFit} model shows it drops to 12\,K at 500 au, leading to a stronger Boltzmann distribution correction factor.

We did not detect the lines \hhdp, \CS, and \soo. We calculated upper limits on disk averaged surface densities for these molecules using Equation \eqref{eq:Nt_thin}, while assuming optically thin emission. The calculations considered 3$\sigma$ values integrated over the line width of N$_2$H$^+$ (coming from Gaussian fit in Figure \ref{fig:int_spectra}). Here, the noise $\sigma = f \times \delta S \times \sqrt{\delta v \times \Delta v}$, where f is the spectral correlation factor of 1.6, $\delta S$ is the root mean squared noise in the flux, $\delta v$ is the channel width and $\Delta v$ is the line width over which noise is calculated. The excitation temperature was fixed at 15 K for this purpose.
The integration area $\Omega$ is considered for a radial region ranging from 150 au to 450 au. The constrained values are tabulated in Table \ref{tab:col_den}.

Our simple LTE approach reports an upper limit on total (summation of both ortho and para species) \ce{H2D+} surface density of $7.48 \times 10^{12}$ cm$^{-2}$, while our \ce{DiskFit} analysis quotes an upper limit on ortho-\ce{H2D+} surface density of $2.1 \times 10^{10}$ cm$^{-2}$. Converting this to a total H$_2$D$^+$ surface density requires an assumption on the ortho-to-para ratio. Under the hypothesis of thermalization, this ratio can be very small since the ortho-H$_2$D$^+$ ground state is 86 K above the para ground state, leading to a huge Boltzmann correction factor (from 70 to 300 for temperatures ranging from 20 down to 15 K respectively). At 15 K, the 3$\sigma$ upper limit on the total \ce{H2D+} surface density, derived from the \ce{DiskFit} ortho-\ce{H2D+} value, is $6.3 \times 10^{12}$ cm$^{-2}$.

Note that the low-level detection of $^{13}$CS (2-1) with NOEMA, a low energy transition reported by \citet{Phuong2021},  implies a surface density of $3 \times 10^{11}$ cm$^{-2}$ for $T_{rot} = 15$ K, below our upper limit.

\begin{deluxetable*}{lcccc}
\tablecaption{Disk averaged surface densities of the detected molecules and upper limits on the undetected molecules under LTE assumption}
\label{tab:col_den}
\tablewidth{\textwidth}
\tablehead{
  \colhead{Species} & \colhead{Transition} & \colhead{surface density, N$_{T}$ (cm$^{-2}$)} & \colhead{Excitation Temperature, T$_{ex}$ (K)} & \colhead{Optical Depth, $\tau$}
}
\startdata
N$_2$H$^+$ & 4-3 & $(1.19 \pm 0.04) \times 10^{12}$ & 12 (fixed) & 0.42 \\
DCO$^+$ & 5-4 & $(1.27 \pm 0.08) \times 10^{11}$ & 16 (fixed) & 0.04 \\
H$_2$D$^+$ & 1(1,0)-1(1,1) & $<7.48 \times 10^{12}$ & 15 (fixed) & -  \\
$^{13}$CS & 8-7 & $<1.12 \times 10^{12}$ & 15 (fixed) & -  \\
SO$_2$& 19(4,16)-19(3,17) & $<2.46 \times 10^{16}$ & 15 (fixed) & - \\
\enddata
\tablecomments{
\newline
a) Surface density error are coming from 16th and 84th percentile of the posterior distribution.\\
b) The upper limits on the disk averaged surface densities of the undetected molecules (H$_2$D$^+$, $^{13}$CS and SO$_2$) are constrained using the  3$\sigma$ upper limit on the line flux as explained in Section \ref{sec:rotational_diagram}.\\
c) In this table, we have reported the upper limit on the total (sum of ortho and para spin-states) \ce{H2D+} surface density. The $3\sigma$ upper limit for o-\ce{H2D+} from \textsc{DiskFit} analysis is $2.1 \times 10^{10}$ cm$^{-2}$.}
\end{deluxetable*}
\section{Astrochemical Modeling} \label{sec:models}
\subsection{Disk Physical Model} \label{subsec:phy_model}
Our physical model is constructed for a radial range of 190 au to 290 au, which primarily includes the ring region around the central stellar system GG Tau A. Only 10-20\% of the total disk mass is contained in the outer disk region, which, therefore, exhibits a lower density than in the ring. Moreover, the outer disk, under the shadow of the dense ring, is very cold \citep{Tang2016, Dutrey2014, Brauer2019} with poor constraints on the physical properties. We chose not to focus our present modelling effort on this outer region, which has a more complex structure \citep{Tang2016,Phuong2020}.\par

In the ring, the radial distribution of the midplane (T$_{mid}$) and the atmospheric temperature (T$_{atm}$) are taken as power laws, constrained in previous studies by \citet{Dutrey2014} and \citet{Guilloteau1999} respectively. They are of the form
\begin{align}
	T_{mid} (r) &= T_{mid, R_{\text{ref}}} \left( \frac{r}{R_{\text{ref}}} \right)^{-q} \label{eq:T_mid}\\
	T_{atm} (r) &= T_{atm, R_{\text{ref}}} \left( \frac{r}{R_{\text{ref}}} \right)^{-q}\label{eq:T_atm}
\end{align}

The values of the parameters can be found in Table \ref{tab:physical_model}. The radial dependency of  scale height, H$_r$ at the disk midplane is calculated assuming hydrostatic equilibrium using the midplane temperature (Equation \ref{eq:T_mid}), the mass of the stellar system, $M_* = 1.37 M_{\odot}$ (rescaled from \citet{Guilloteau1999} considering a distance of 150 pc), the mean molecular weight of the gas per H nuclei $\mu = 2.37$ and the atomic mass unit, $m_H$, as follows
\begin{align}
	H_r= \sqrt{\frac{k_B T_{mid} r^3}{\mu \, m_H \, G M_*}} \label{eq:H}
\end{align}
Here $k_B$ and $G$ are Boltzmann and Gravitational constants, respectively. We constructed the 1D vertical structure at each radial point until 4H$_r$ height. The vertical temperature structure is adopted from \citet{Dartois2003} and \citet{Williams2014} given by,
\begin{align}
    T(z) = T_{mid} + (T_{atm} - T_{mid})\left[ sin \left(\frac{\pi z}{2 z_q}\right)\right]^{2\sigma}
\end{align}

$T_{mid}$ and $T_{atm}$ are defined at the midplane (z = 0) and the upper end (z = z$_q$) of the atmosphere assuming Equations \ref{eq:T_mid} and \ref{eq:T_atm}, respectively. $z_q$ is set at 4H$_{r}$ in our case. The parameter $\sigma$ denotes the stiffness of the vertical temperature distribution. As pointed out by \citet{Tang2016}, the inner edge of the GG Tau ring obstructs a substantial amount of starlight, minimizing the vertical temperature variation within. Therefore, we have chosen a value of 0.5 for $\sigma$. We have kept the dust temperature equal to the gas temperature, as dust and gas are relatively well coupled for such densities ($>10^8$ cm$^{-3}$). \par

The gas surface density, $\Sigma_g(r)$ [g cm$^{-2}$] is parameterized as a power law profile suggested by \citet{Lynden1974}, given by:
\begin{align}
	\Sigma_g(r) = \Sigma_{g, R_{\text{ref}}} \left( \frac{r}{R_{\text{ref}}} \right)^{-\gamma} \label{eq:sigmag_r}  
\end{align}

The power law index $\gamma$ is taken to be 1.4 following \citet{Dutrey2014}. The gas surface density at the reference radius $R_{\text{ref}}$, $\Sigma_{g, R_{\text{ref}}}$ is calculated considering the mass contained in the circumtertiary ring, $M_{g} = 0.13 M_{\odot}$ \citep{Guilloteau1999, Andrews2014}. The calculations are as follows:
\begin{align}
	M_g = 2\pi \int_{R_{in}}^{R_{out}} \Sigma_g(r)rdr\label{eq:Mg}
\end{align}

Substituting Equation \ref{eq:sigmag_r} in \ref{eq:Mg}, for $R_{\text{ref}}=214$ au, we get:
\begin{align}
	\Sigma_{g, R_{\text{ref}}} = \frac{(2-\gamma) M_g}{2 \pi R_{\text{ref}}^\gamma} \left[ R_{\text{out}}^{2-\gamma} - R_{\text{in}}^{2-\gamma}\right]^{-1} = 9.7 \text{ g cm$^{-2}$}
\end{align}

We then calculate the \ce{H2} number density [cm$^{-3}$] at the midplane with,
\begin{align}
    n_{H_2,midplane} = \frac{\Sigma_g(r)}{\mu \,m_H\,H_r\sqrt{2\pi}}
\end{align}
Using the midplane density as the lower limit, we integrate the equation of hydrostatic equilibrium to obtain the vertical density structure \citep{Reboussin2015}:
\begin{align}
    \frac{\partial \ln{n_{H_2}(z)} }{\partial z} = - \left[ \left( \frac{G M_* z}{r^3} \right) \left( \frac{\mu m_H}{k_B T} \right) + \frac{\partial \ln{T}}{\partial z}\right]
\end{align}

The local visual extinction is determined from hydrostatic density structure, assuming a conversion factor of $(A_v/N_H)_0 = 6.25 \times 10^{-22}$ \citep{Wagenblast1989} where $N_H = 2N_{H_2}$ is the vertical hydrogen surface density. We assume $N_H$, and consequently $A_v$ are zero above 4H$_r$. To account for the impact of grain size ($r_d$) and dust-to-gas mass ratio ($\varepsilon$), we scale the conversion factor as follows \citep{Wakelam2019}:
\begin{align}
    \frac{A_v}{N_H} = \left( \frac{A_v}{N_H} \right)_0 \frac{\varepsilon}{10^{-2}} \frac{10^{-5}}{r_d(cm)}
\end{align}

The stellar UV flux factor $f_{\text{UV}}$ follows the inverse squared law of distance and acts on the top of the disk atmosphere, only half of which is scattered towards the midplane, the rest being absorbed or scattered upwards \citep{Wakelam2016}:

\begin{align}
    f_{\text{UV}} (r) = \frac{1}{2}\frac{f_{\text{UV}, R_{\text{ref}}}}{\left( \frac{r}{R_{\text{ref}}}\right)^2 + \left( \frac{4H_r}{R_{\text{ref}}} \right)^2}
\end{align}
$f_{\text{UV}, R_{\text{ref}}}$ is defined at the surface of the disk in the units of interstellar DRAINE field spectrum, $\chi_0$ \citep{Drain1978} and reported in Table \ref{tab:physical_model}. H$_r$ is calculated using Equation \ref{eq:H}. The vertical distribution of UV flux is obtained by multiplying $f_{UV}$ with $e^{-A_v/1.086}$, where $A_v$ is the visual extinction distribution \citep{Du2014}.\par

\begin{figure*}
    \centering
    \includegraphics[width=\textwidth]{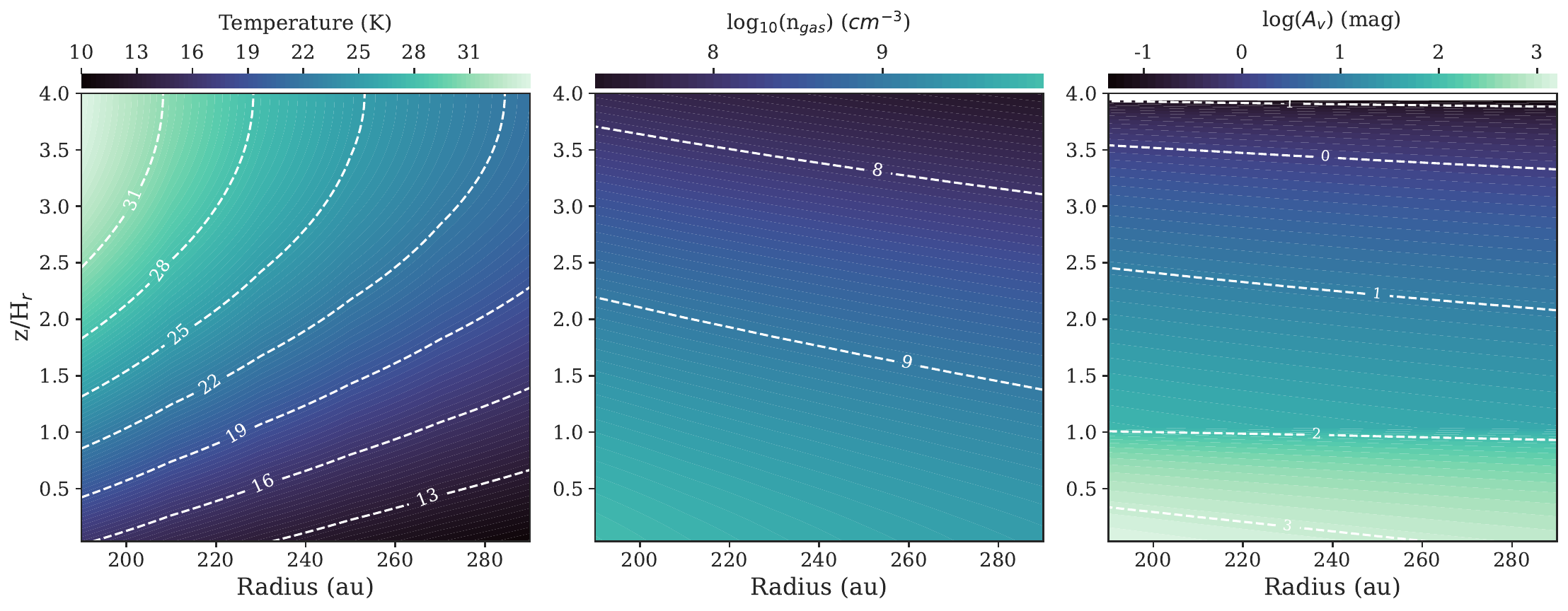}
    \caption{Physical structure of the ring region in our model derived using the parameters tabulated in Table \ref{tab:physical_model}. The left one indicates the temperature distribution, the middle one is the gas density distribution in the log scale, and the one on the right is visual extinction in the log scale. The description of the model is explained in Section \ref{sec:models}.}
    \label{fig:best_fit_structure}.
\end{figure*}

In our models, we have not implemented grain growth; instead, we considered a uniform grain size of 0.1$\mu m$ \citep{Phuong2018} settling below 1H$_r$.
To incorporate dust settling, we adjust the $\varepsilon$ to $10^{-3}$ above 1H$_r$, while maintaining the standard $\varepsilon$ of $10^{-2}$ below 1H$_r$. Figure \ref{fig:best_fit_structure} illustrates the corresponding structure of the best physical model.

\begin{deluxetable}{lcc}[!t]
\tablecaption{Parameter prescription of the best-fit model considered\label{tab:physical_model}}
\tablehead{
\colhead{Parameter Description} & \colhead{Values} & \colhead{Units}
}
\startdata
\cutinhead{Fixed Parameters}
Stellar mass: $M_*$ & 1.37 & M$_\odot$\\
Ring mass: $M_g$ & 0.13$^a$ & M$_\odot$\\
Ring inner radius: $R_{\text{in}}$ & 193 & au\\
Ring outer radius: $R_{\text{out}}$ & 285 & au\\
Reference radius: $R_{\text{ref}}$ & 214 & au \\
Midplane temperature at $R_{\text{ref}}$: $T_{\text{mid, R$_{\text{ref}}$}}$ & 14 & K\\
Atmospheric temperature at $R_{\text{ref}}$: $T_{\text{atm, R$_{\text{ref}}$}}$ & 30 & K\\
Surface density at $R_{\text{ref}}$: $\Sigma_{g,\text{ref}}$ & 9.7 & g cm$^{-2}$\\
Temperature power-law index: $q$ & 1.1 & \\
Temperature vertical gradient index: $\sigma$ & 0.5 & \\
Surface density power law index: $\gamma$ & 1.4 & \\
Grain size: $r_d$  & 0.1 & $\mu m$ \\
\cutinhead{Adjusted Parameters}
UV Flux: $f_{\text{UV}, R_{\text{ref}}}$  & 375 & Draine's unit ($\chi_0$)\\
Cosmic ray ionization, $\zeta_{CR}$ & $10^{-18}$ & s$^{-1}$\\
Dust settling & Yes & \\
C/O ratio & 1.0 &\\
Age of the parent molecular cloud  & $10^6$ & yr\\
Age of the disk & $10^6$ & yr
\enddata
\tablecomments{
All the fixed values are taken from \citet{Guilloteau1999,Dutrey2014} and $^a$\citet{Andrews2014} and rescaled considering a distance of 150 pc, except $\sigma$ and $\Sigma_{\text{ref}}$. $\sigma$ is expected to be small as the inner edge of the ring casts a shadow on the rest of the structure \citep{Tang2016}. $\Sigma_{\text{ref}}$ is calculated using Equation \ref{eq:sigmag_r} assuming 0.13$M_{\odot}$ of total disk mass is contained in the ring region.
}
\end{deluxetable}

\subsection{Chemical Model and Network}\label{subsec:astrochem_model}
The gas-grain astrochemical model \textsc{Dnautilus}, introduced in \citet{Majumdar2017}, is used to compute the 1+1D chemical composition in the ring of GG Tau A by incorporating all the updates from \textsc{Dnautilus 2.0} outlined in \citet{Kotomi2024}. \textsc{Dnautilus 2.0} has the capability to investigate deuterium fractionation in both two-phase (gas and grain surface) and three-phase (gas, grain surface, and grain bulk) modes. \textsc{Dnautilus 2.0} includes 1606 gas species, 1472 grain-surface species, and 1472 grain-mantle species, connected by 83,715 gas-phase reactions, 10,967 reactions on grain surfaces, and 9431 reactions in the grain mantles.

Although including spin-state chemistry of hydrogenated species like \ce{H2}, \ce{H2+}, \ce{H3+}, and their isotopologues would have allowed us to directly investigate \ce{o-H2D+} chemistry, we chose not to do so. The primary reason is that the newly proposed state-to-species and species-to-state reaction rates for the \ce{H3+ + H2} systems by \citet{Sipila2017} vary depending on density and temperature regimes. Applying them to a protoplanetary disk would require rigorous benchmarking and testing across various star-forming environments, and such extensive validation is beyond the scope of this paper.

\subsection{Grid of Models}
\begin{figure*}
    \centering
    \includegraphics[width=0.9\textwidth]{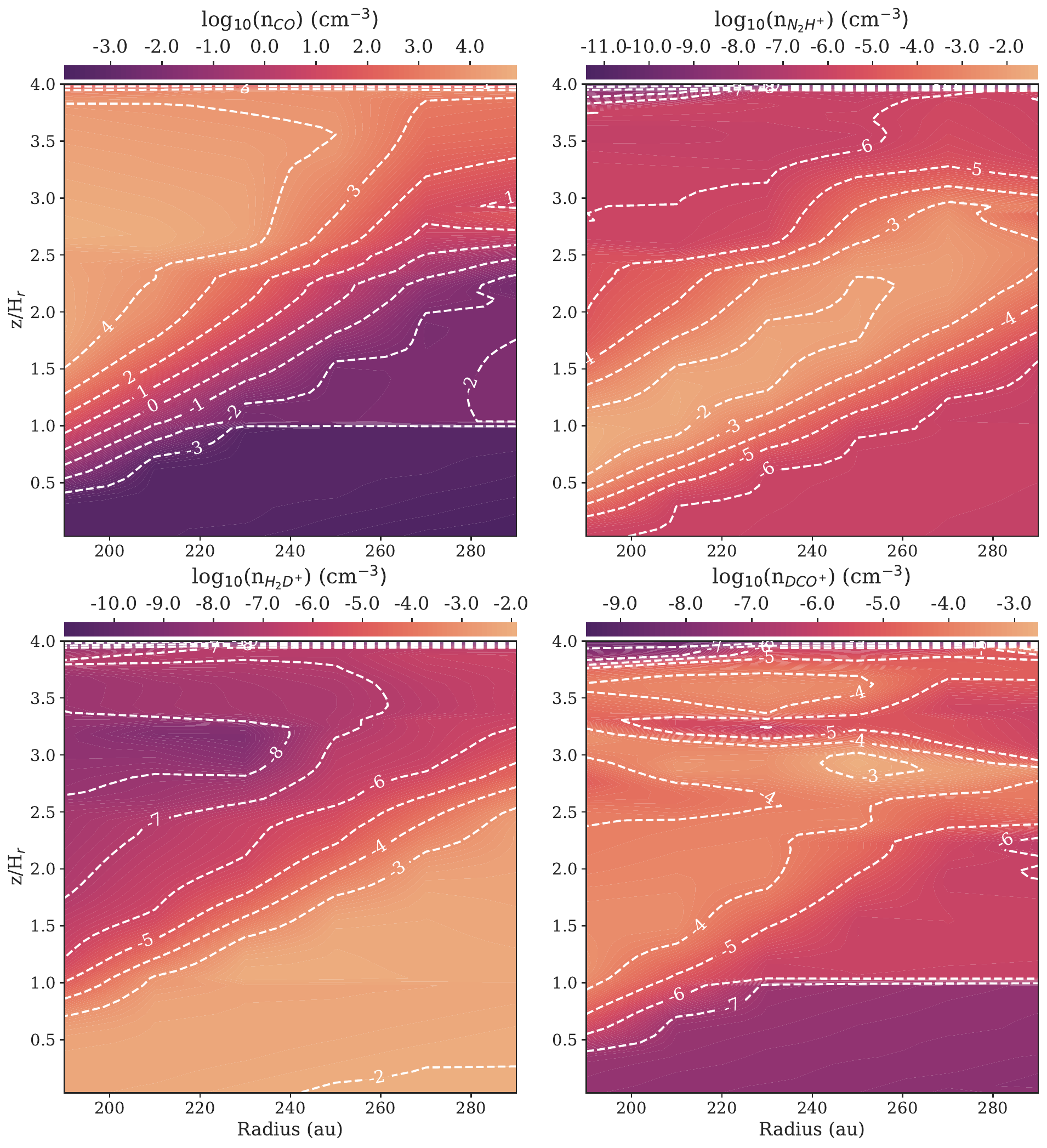}
    \caption{Number density distributions (log-scale) from our best-fit model for \ce{CO}, N$_2$H$^+$, H$_2$D$^+$ and DCO$^+$ from left to right and top to bottom. We have only modelled the ring region around GG Tau A, spanning a radial distance of 193 au to 285 au from the central stellar system. The model disk atmosphere extends up to 4 scale heights.}
    \label{fig:modelled_abundances}
\end{figure*}

To find the model that best represents our observationally constrained radial surface density distribution of \ce{N2H+}, \ce{DCO+}, and the upper limits on \ce{H2D+}, \ce{SO2}, and $^{13}$CS \citep[including the constraint from the $J=2-1$ transition by][]{Phuong2021}, we explored a grid of models utilizing the least squares method (refer to Appendix \ref{app:chemical_model}). These models span a range of initial carbon-to-oxygen (C/O) ratios, cosmic ray ionization (CRI) rates and UV flux values (see Table \ref{tab:dnautilus_parameters} in Appendix \ref{app:chemical_model}). The model with the least disagreement with observations ($\chi^2_{\text{red}}$ = 2.78) is an inheritance disk model with initial abundances characterized from a starless dense molecular cloud, illuminated by stellar UV radiation of 375$\chi_0$ at a reference radius of 214 au. First, we allowed the molecular cloud to evolve chemically for $10^6$ years from initial atomic abundances (Table \ref{tab:initial_abun}) with typical physical conditions such as gas and dust temperature of 10 K, a total gas density of $2\times10^4$ cm$^{-3}$, a visual extinction of 15 mag and cosmic ray ionization rate of $1.3 \times 10^{-17}$ s$^{-1}$. We then computed the temporal evolution of chemistry for the inherited material from the cloud in the ring region around GG Tau A for $10^6$ years. The assumption that protoplanetary disks form from a cold, starless, dense molecular cloud with the mentioned physical conditions is based on detailed discussions from \citet{Wakelam2019}, without taking into account the chemical evolution between the cold core phase and the protoplanetary disk itself \citep{Drozdovskaya2016}. It is crucial to highlight that in order to reproduce the observed surface densities in our disk simulation, a cosmic ray ionization rate ($\zeta_{\text{CR}}$) of about $10^{-18}$ s$^{-1}$ and an initial C/O ratio of 1.0 are necessary. Table \ref{tab:physical_model} presents the parameters describing the best-fit model. Figure \ref{fig:modelled_abundances} showcases the modelled number density distributions within the ring region.

\begin{deluxetable}{lcc}
\tablecaption{Initial abundances used in our cloud model\label{tab:initial_abun}}
\tablewidth{\columnwidth}
\tablehead{
\colhead{Element} & \colhead{Abundance relative to H} & \colhead{References}
}
\startdata
\ce{H2}  & $0.5$ &  \\
He  & $9.00 \times 10^{-2}$ & 1 \\
N   & $6.20 \times 10^{-5}$ & 2 \\
O   & $(1.40 - 3.3) \times 10^{-4}{}^{(a)}$ & 3 \\
\ce{C+}  & $1.70 \times 10^{-4}$ & 2 \\
\ce{S+}  & $8.00 \times 10^{-8}$ & 4 \\
\ce{Si+} & $8.00 \times 10^{-9}$ & 4 \\
\ce{Fe+} & $3.00 \times 10^{-9}$ & 4 \\
\ce{Na+} & $2.00 \times 10^{-9}$ & 4 \\
\ce{Mg+} & $7.00 \times 10^{-9}$ & 4 \\
\ce{P+}  & $2.00 \times 10^{-10}$ & 4 \\
\ce{Cl+} & $1.00 \times 10^{-9}$ & 4 \\
F   & $6.68 \times 10^{-9}$ & 5 \\
HD  & $1.60 \times 10^{-5}$ & 6 \\
\enddata
\tablerefs{ (1) \citet{Wakelam2008}; (2) \citet{Jenkins2009}; (3) \citet{Reboussin2015} (4) Low metal abundances from \citet{Graedel1982}; (5) Depleted value from \citet{Neufeld2005}; (6) \citet{Majumdar2017}. (a) O initial abundance is adjusted for C/O ratio ranging from 0.5 to 1.2} 
\end{deluxetable}

\begin{figure*}
    \centering
    \includegraphics[width=\linewidth]{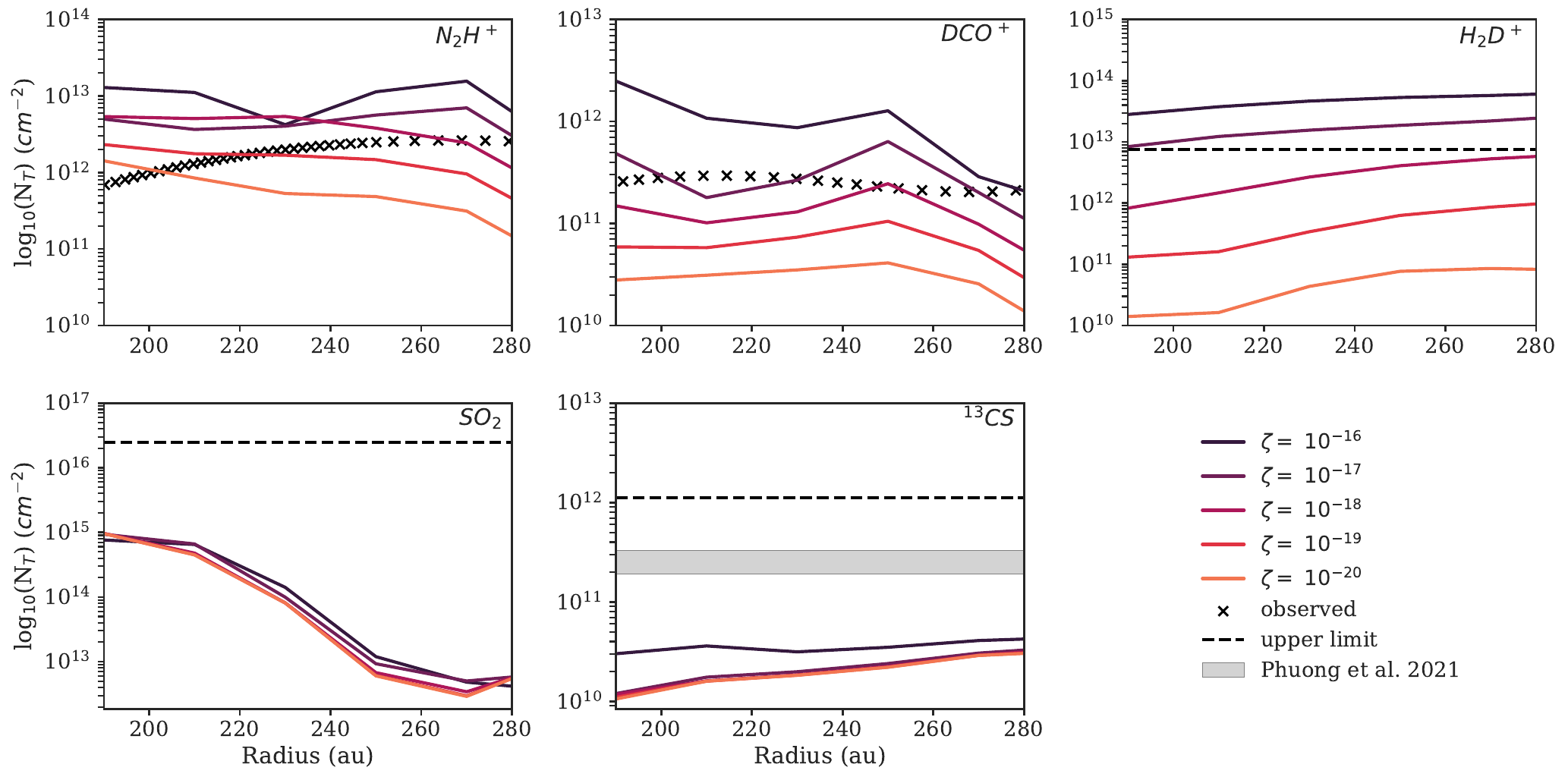}
    \caption{Comparison between models with different CR rates ($\zeta$) represented by the solid coloured lines in the figure. Here, the `x' points represent the observationally contained surface density values from our \textsc{DiskFit} analysis. The dashed horizontal lines represent observationally constrained upper limits on surface density for the undetected molecules. The grey horizontal patch denotes the surface density constraint from a different transition of $^{13}$CS (J=2-1) by \citet{Phuong2021}. Best model: CRI = $10^{-18}$ s$^{-1}$ ($\chi^2_{\text{red}} = 5.01$). Here, C/O = 0.7 and the rest of the parameters are set to the best-fit parameters. The upper limit on $^{13}$CS surface density is calculated from the CS surface density upper limit by assuming N$_{\text{CS}}$/N$_{^{13}\text{CS}}$ = 100 \citep{Phuong2021}. }
    \label{fig:comp_cr}
\end{figure*}

\begin{figure*}
    \centering
    \includegraphics[width=\linewidth]{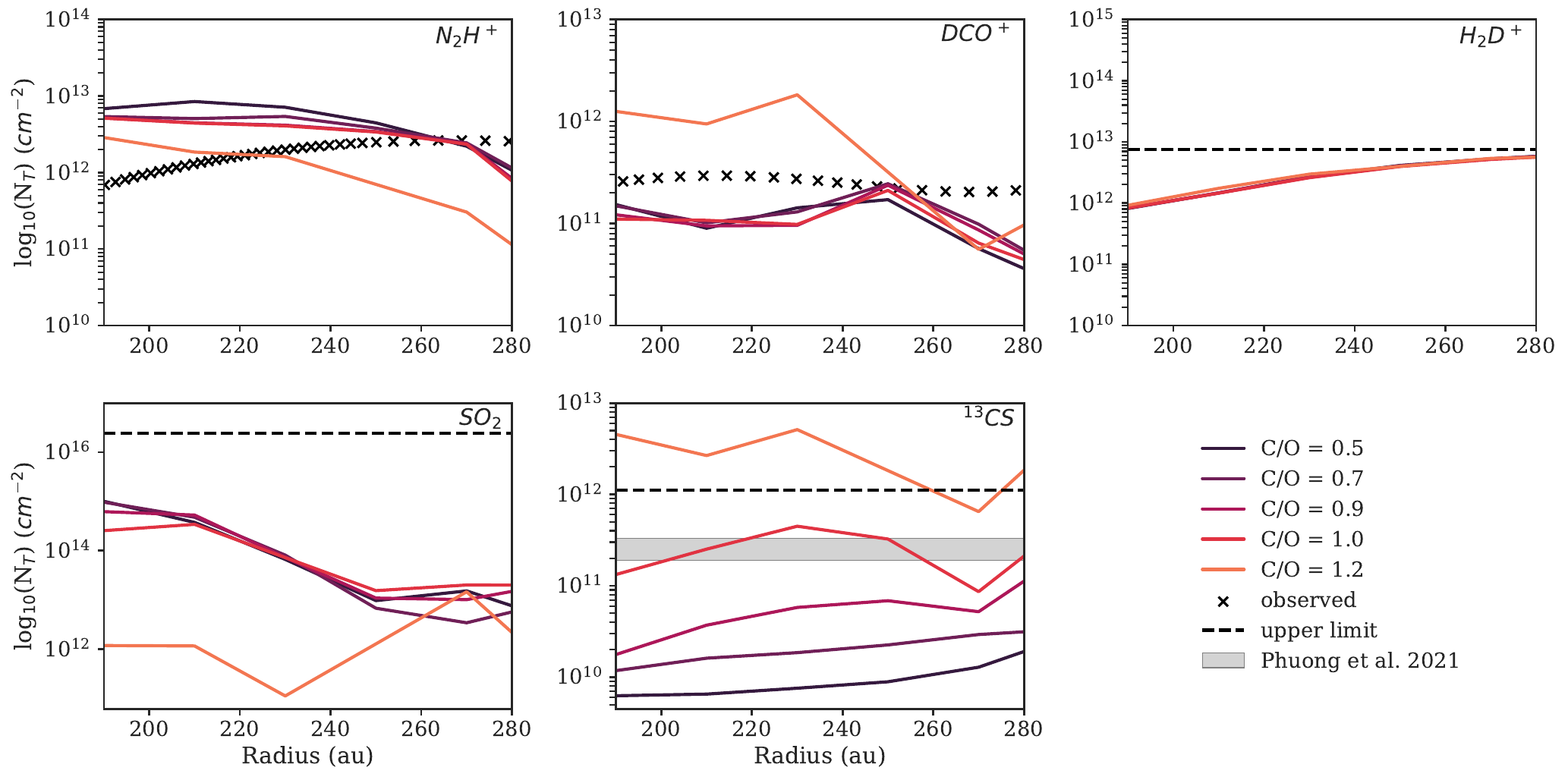}
    \caption{Comparison between models with different initial C/O ratios represented by the solid coloured lines in the figure. The rest of the denotations are the same as Figure \ref{fig:comp_cr}. Best model: C/O = 1.0 ($\chi^2_{\text{red}} = 2.78$). All the parameters other than C/O are set to the best-fit parameters. The upper limit on $^{13}$CS surface density is calculated from the CS surface density upper limit by assuming N$_{\text{CS}}$/N$_{^{13}\text{CS}}$ = 100 \citep{Phuong2021}.}
    \label{fig:comp_co}
\end{figure*}

\section{Discussion} \label{sec:discussions}
\subsection{Comparison with Previous Observations}
\hhdp\ has been observed in multiple dense starless cores and prestellar cores \citep[see for a review]{Ceccarelli2014}.
\citet{Caselli2008} reports o-\ce{H2D+} surface density $\sim 10^{12} - 10^{13}$ cm$^{-2}$ in regions characterized by gas density $\sim 10^{5} - 10^{6}$ cm$^{-3}$ and kinetic temperatures around 10 K, with most prominent detections in the densest and centrally concentrated cores. However, no clear detection has been obtained in protoplanetary disks. \citet{Chapillon2011} derived $3 \sigma$ upper limits on o-\ce{H2D+} surface density for the T Tauri disks DM Tau and TW Hya. Depending on the physical model, DM Tau's upper limit lies between $4.5 \times 10^{11}$ and $1.9 \times 10^{12}$ cm$^{-2}$ and for TW Hya, it is between $9.0 \times 10^{11}$ and $1.4 \times 10^{12}$ cm$^{-2}$. Our upper limit for o-\ce{H2D+} in GG Tau is $2.1 \times 10^{10}$ cm$^{-2}$. Note that \citet{Chapillon2011}'s upper limits are from single-dish observations (1$\sigma$ noise level 0.33 Jy km s$^{-1}$). We have achieved a much higher sensitivity (1$\sigma$ noise level 0.019 Jy km s$^{-1}$), making our observation the best available constraint on the upper limit of o-\ce{H2D+} surface density in any disk to date.

The \ce{N2H+} (3-2) transition has been observed in all five disks (three around T Tauri stars: IM Lup, GM Aur, and AS 209, and two around Herbig Ae stars: HD 163296 and MWC 480) as part of the Molecules with ALMA at Planet-forming Scales (MAPS) program \citep{MAPSX}. Despite the diverse continuum structures of the disks with multiple rings and gaps, \ce{N2H+} emission showcases a ring morphology (Figure 4 in \citet{MAPSX}). Our \ce{N2H+} emission is also characterized by a single-ring structure (Figure \ref{fig:radial_intensity}). \citet{Qi2013} has previously reported \ce{N2H+} detection towards TW Hya, and the single-ring emission structure is also evident there. The \ce{N2H+} ring morphology across different disks can be attributed to radial and vertical temperature variations \citep{Qi2013}. In the MAPS survey by \citet{MAPSX}, two different \ce{N2H+} surface density values are derived by fixing the excitation temperature, $T_{\text{ex}}$, to a) the observationally constrained CO freeze-out temperature $\sim 20$ K and b) the midplane temperatures, $T_{\text{mid}}$, of the respective disks. For the T Tauri disks (IM Lup, GM Aur, and AS 209), the \ce{N2H+} surface density peaks around 10$^{13}$ cm$^{-2}$, while for the Herbig Ae disks (HD 163296 and MWC 480), it reaches up to 10$^{12}$ cm$^{-2}$. Interestingly, GG Tau, despite being a T Tauri disk, exhibits a \ce{N2H+} surface density of approximately 10$^{12}$ cm$^{-2}$. The surface density in TW Hya reported by \citet{Qi2013} ranges from $4 \times 10^{12}$ to $2 \times 10^{15}$ cm$^{-2}$ for different assumed physical structures. \citet{MAPSX} indicates that temperatures as low as 12 K in the \ce{N2H+} emitting region can also explain the observed flux of the 3-2 transition in the protoplanetary disks (e.g., GM Aur, HD 163296, MWC 480) under the LTE assumption. Previously, in \citet{Phuong2021}'s study, \ce{N2H+} in GG Tau A was assumed to be arising from the same layer as CO, following a radial temperature variation, $T(r) = 20 (r/250 \text{ au})^{-1}$ K. Our direct measure of $T(r) = 12$\,K for the \ce{N2H+} emitting region (Table \ref{tab:diskfitabun}), based on multi-line excitation analysis, suggests that we are tracing a region much closer to the midplane than initially assumed.

A ring-like radial distribution for \ce{DCO+} has been commonly observed in multiple disks, including the TW Hya, DM Tau, AS 209, V4046 Sgr, MWC 480 and HD 163296 \citep{Qi2008,Mathews2013,Teague2015,Huang2017} and is facilitated by reaction between \ce{H2D+} and \ce{CO} (e.g. \citet{Willacy2007}). Contrary to the common scenario, we observe partially overlapping double-ring radial distribution of \ce{DCO+} emission in GG Tau A (refer to Figure \ref{fig:radial_intensity}), which can likely be attributed to different chemical effects in the ring and the outer disk regions owing to the difference in densities between the two regions. This is discussed in detail in Sections \ref{subsec:chem_ring} and \ref{subsec:chem_outerdisk}. Notably, \citet{Oberg2015a} observes double \ce{DCO+} rings in the disk IM Lup. They found that \ce{DCO+} abundance first decreases radially due to CO depletion, then another ring forms due to non-thermal desorption of CO in low surface density outer regions as a result of higher UV penetration. \citet{Phuong2018,Phuong2021} have previously assumed that \ce{DCO+} emission originates from the same region as \ce{N2H+} and \ce{CO} characterized by radial temperature, $T(r) = 20 (r/ 250 \text{ au} )^{-1}$ K in GG Tau A. In this work, we found that \ce{DCO+} emission region is best represented by $T(r) = 16 (r/250 \text{ au})^{-0.39}$ K (Table \ref{tab:diskfitabun}). Hence, we expect that \ce{DCO+} emission region lies vertically between \ce{CO} emitting region and \ce{N2H+} emitting region and closer to the midplane than previously thought.

\subsection{Chemical Trends in the Ring Region} \label{subsec:chem_ring}
We present a sub-set of chemical models for a selected set of input parameters in Figures \ref{fig:comp_cr} and \ref{fig:comp_co} from the full model grids presented in Table \ref{tab:dnautilus_parameters}. Only the CRI rates and the C/O ratio were found to have any significant impact on the modelled surface densities of the targeted molecules. Other parameters cannot be constrained by the current data.

\paragraph{Cosmic Ray Ionization}
Figure \ref{fig:comp_cr} shows the variation of modelled abundances with CRI rates while keeping the C/O ratio constant at 0.7, and the remaining parameters are set to the best-fit values. \ce{N2H+}, \ce{DCO+}, and \ce{H2D+} abundances are quite sensitive to CRI rates, while UV flux values have no impact whatsoever. This is expected because these molecules probe a high-density, low-temperature, CO-depleted region near the mid-plane where UV penetration is minimal. Hence, cosmic rays are the primary ionization source for this cold molecular chemistry. However, the CRI rate has barely any impact on S-bearing species. It is worth noting that the standard interstellar CRI rate of $10^{-17}$ s$^{-1}$ predicts the \ce{H2D+} surface density above the upper limit and a sub-interstellar CRI rate of $10^{-18}$ s$^{-1}$ explains the observationally constrained surface densities the best. This low CRI rate can be explained by a rise in the number of collisions experienced by CR particles due to the greater effective surface density as they gyrate towards the disk midplane \citep{Padovani2013b}. The attenuation is especially prominent in our case, as our chemistry focuses on high-density near midplane regions. \citet{Cleeves2015} have previously inferred a very low CRI  ($10^{-19}$ s$^{-1}$) in the protoplanetary disk TW Hya, which they suggest is due to an exclusion by a young stellar wind. Our current astrochemical model only includes a constant CR ionization rate across the disk. The inclusion of short-lived radionuclides \citep{Cleeves2013} and vertical distribution of CRI rates \citep{Padovani2018} may provide a better picture.

\paragraph{C/O ratio}
We have also modelled abundances of our targeted molecules for different elemental C/O ratios from 0.5 to 1.2 by changing initial O abundances, as shown in Figure \ref{fig:comp_co}. Changing the C/O ratios from 0.5 to 1.0 has decreased the \ce{N2H+} surface density by a factor of approximately 2.5, while the \ce{DCO+} surface density increases by at most a factor of 1-2. However,
 when the C/O ratio was further increased to 1.2, changes in the surface densities of \ce{N2H+} and \ce{DCO+} became significant (by a factor of approximately 10). \ce{H2D+} surface density remains unaffected for any change in C/O ratio from 0.5 to 1.2.
 The chemistry of \ce{^{13}CS} and \ce{SO2}, for which we obtained the upper limit from our observations, is quite sensitive to changes in the C/O ratio. The general trend is that O-poor chemistry favours the formation of the C-S bond, and for \ce{SO2}, it is the other way around
\citep[see discussions by][for further details]{Semenov+2018,LeGal2021}.
Our observations are best represented by a CRI rate of $10^{-18}$ s$^{-1}$ and an elemental C/O ratio of 1.0. At the commencement of the evolution of a molecular cloud, a significant amount of important elements, including carbon (C) and oxygen (O), exist in a refractory state, which is expected to shape the chemical budget of the protoplanetary disk formed from that cloud. Based on \citet{Boogert2015}'s observation on ice abundances, \citet{Oberg2021} reports that about 40\% O chemical budget was unaccounted for at the onset of star and planet formation, leading to a poor constraint on initial C and O gas phase abundances. Our elemental C/O ratio of 1.0 suggests an oxygen-depleted environment in the dense ring. 
Similar inferences of high C/O ratios ($0.9-2.0$) were also found in more typical protoplanetary disks  \citep[e.g.][]{Ruaud2022, LeGal2021}, which may have consequences on the formation of planets in the inner regions.

\paragraph{Spatial Distributions}
Figure \ref{fig:modelled_abundances} represents the molecular number density distribution for our best model. We see a \ce{DCO+} distribution in those regions where CO is available in the gas phase in small amounts. Observationally, \ce{DCO+} follows a ring-like emission distribution in the ring region (refer to Figures \ref{fig:radial_intensity} and \ref{fig:kepler-n2h}). In our modelled abundances, \ce{DCO+} is primarily peaking up in temperature regions representative of the warm molecular layer (2.5- 3.5H$_r$), somewhat showcasing the ring-like distribution. One of the dominant ions in the warm molecular layer is \ce{HCO+}. It essentially forms through the protonation of \ce{CO} and, in turn, produces DCO$^+$ through isotope exchange. The major reaction for DCO$^+$ production contributing to our observation above 2.5H$_r$ is \ce{HCO+ + D -> DCO+ + H}. We also see some \ce{DCO+} arising below the warm molecular layer ($\sim 0.5 - 2.5$ H$_r$) produced through a different channel, \ce{CO + N2D+ -> N2 + DCO+}. Our \textsc{DiskFit} analysis locates the peak of the \ce{DCO+} ring at radial distance $210 \pm 50$ au (R$_1$ in Table \ref{tab:diskfitabun}), and this can be attributed to the fact that both the \ce{DCO+} formation channels are contributing to \ce{DCO+} abundances in this radial range. Near the disk midplane, \ce{DCO+} starts depleting because its parent molecule, \ce{CO}, efficiently freezes onto the grain surfaces owing to high density and low temperature.

\ce{N2H+} modelled distribution lies just below the \ce{DCO+} layer, with its number density peaking up in the regions where \ce{DCO+} starts depleting. Our observationally constrained representative temperatures of 16 K and 12 K for \ce{DCO+} and \ce{N2H+} emitting regions, respectively (Table \ref{tab:diskfitabun}), suggest the same. N$_2$H$^+$ is abundant in the midplane at the inner edge of the ring as well as in the warm molecular layer (up to 3.0H$_r$) at the outer edge of the ring. All throughout the ring region, the major formation pathway of N$_2$H$^+$ is the protonation of \ce{N2}, i.e., \ce{N2 + H3+ -> H2 + N2H+}. So, the chemistry is mainly controlled by gas phase abundances of \ce{CO}, \ce{N2}, \ce{HD} and \ce{H3+}. As long as \ce{CO} remains in the gas phase, it destroys \ce{N2H+} to form \ce{HCO+}. When CO starts depleting, \ce{N2H+} start increasing, as \ce{N2} is still in the gas phase owing to its lower binding energy with respect to CO, forming \ce{N2H+} and there is less destruction of \ce{N2H+} because of the reduced CO presence. Our model predicts a decreasing \ce{N2H+} surface density with increasing radius, contrary to the observed increase within the continuum ring. This discrepancy arises primarily from the model's overestimation of \ce{N2} depletion from the gas phase. A similar trend was observed in the protoplanetary disk around DM Tau, where \citet{Wakelam2019} conducted sensitivity analyses on various model parameters. One limitation of our model is the absence of a grain growth mechanism and the assumption that grains are perfectly spherical. \citet{Kataoka2013} proposed a dust growth mechanism in protoplanetary disks, which would create inhomogeneities in the dust distribution and potentially allow more UV penetration towards the disk's mid-plane. UV penetration, in turn, facilitates photo-desorption of \ce{N2}. Chemically, non-spherical grains would have a larger cross-sectional area for collisions with gas-phase species and a larger surface area, increasing the likelihood of \ce{N2} desorption. Additionally, our models do not include X-ray chemistry, which could also affect the abundance of \ce{N2H+} \citep{Teague2015, Cleeves2015}.

Our astrochemical model suggests that \ce{H2D+} is abundant in the midplane region, situated just below the \ce{N2H+} layer. The major formation pathway predicted by our model is \ce{H3+ + HD -> H2D+ + H2}. As \ce{N2} begins to freeze onto grain surfaces, more \ce{H3+} become available for \ce{HD} to capture, and we see a rise in \ce{H2D+} abundances and a fall in \ce{N2H+} abundances.

We did not detect \ce{H2D+}. In our effort to replicate the observed surface density for the detected molecules, we realized that a lower cosmic ray ionization rate may provide the best explanation for our observations. \ce{H2D+} abundances are regulated by the availability of \ce{H3+}, which in turn is dependent on the rate of ionization of \ce{H2}. A lower CRI rate results in reduced \ce{H2+} production (\ce{H2 + CR -> H2+ + e-}), leading to a lower abundance of \ce{H3+} (\ce{H2 + H2+ -> H3+ + H }). Hence, a lower ionization rate at the midplane, owing to the very high density in the ring region, may explain the non-detection of \hhdp. Indeed, our model using a CR rate ($\zeta_{CR}$) of $10^{-18}$ s$^{-1}$ predicts a \ce{H2D+} surface density of the order of $\sim 10^{12}$ cm$^{-2}$ close to the current upper limit reported in Table \ref{tab:col_den}. This limit suggests we are close to detection, and increasing the observation time should assure our chance of detection for \ce{H2D+}. However, the upper limit on total \ce{H2D+} is based on the assumption of thermalized ortho-to-para ratio (LTE), with a high correction Boltzmann factor (see Section \ref{sec:surfden_results}).
While thermalization is facilitated by the very high mid-plane density, spin-dependent reaction rates may lead to a different prediction \citep[e.g.][]{Chapillon2011}.

\subsection{Possible Chemistry in the Outer Disk}\label{subsec:chem_outerdisk}
\nnhp\ and \dcop\ emission are also spatially extended beyond the ring, reaching up to a radial distance of approximately 550 au (Figures \ref{fig:mo_maps}, \ref{fig:radial_intensity}). The velocity-integrated radial profiles of \nnhp\ and \dcop\ (Figure \ref{fig:radial_intensity}) indicate that the former is decreasing as we go radially outward in the outer disk, while the latter shows another peak at around $\sim$ 375 au.\par

We have not modelled the outer disk due to its complex structure \citep{Tang2016, Phuong2020}. However, based on our current understanding, the outer disk is characterized by lower density and colder temperatures as compared to the ring (refer to Section \ref{subsec:phy_model}). Lower densities outside the dense ring allow higher CR penetration, leading to non-thermal desorption of CO and \ce{N2}. Consequently, an increased presence of CO and \ce{N2} in the gas phase is anticipated in the outer disk, suggesting that the primary pathway for DCO$^+$ production would be through \ce{H2D+ + CO} \citep{Phuong2018}. On the other hand, \ce{N2H+} formed through protonation of \ce{N2} will be readily destroyed in the presence of \ce{CO}, which will cause a rapid decrease in the \ce{N2H+} abundances as seen in Figure \ref{fig:radial_intensity}.

\subsection{Temperature Trends in the Ring Region}

The \textsc{DiskFit} analysis of our observations of \nnhp\ and \dcop\ along with NOEMA observations N$_2$H$^+$ (1-0) and DCO$^+$ (1-0, 3-2) by \citet{Phuong2021,Phuong2018} indicate that N$_2$H$^+$  and DCO$^+$ are best represented by average temperatures of 12 K and 16 K (Table \ref{tab:diskfitabun}) respectively. These rotation temperatures are consistent with N$_2$H$^+$ being located closer to the cold disk mid-plane than DCO$^+$ and a fortiori, CO and its isotopologues previously studied by \cite{Phuong2020}.\par

However, in our chemical model, while the distribution of \ce{N2H+} and \ce{DCO+} extends to regions with temperatures of 12 and 16 K, their density-weighted average temperatures are 21 K and 24 K, respectively. The discrepancy with observed temperatures can be attributed to our limited understanding of the physical structure of the ring region and the simplifications applied in the model to derive the physical structure. DCO$^+$ abundances are sensitive to grain size, and the dust-to-gas mass ratio significantly influences gas-phase molecular abundances. \citet{Gavino2021, Gavino2023} recently investigated the impact of the dependence of the grain size on their temperature. They found that large grains exhibit lower temperatures than smaller ones, with an effect on chemistry that cannot be properly represented by a single grain size. This can influence the structure of the CO distribution (with more CO near the mid-plane) and, consequently, the gas-phase abundances of DCO$^+$ and N$_2$H$^+$ in the densest, coldest regions.
\section{Conclusions} \label{sec:conclusion}
In this paper, we have presented high-sensitivity ALMA band 7 observations of the circumstellar disk around the triple star system GG Tau A that highlights CO-regulated cold molecular chemistry in the shielded high-density regions near the disk midplane. Our key findings and conclusions are summarized below:
\begin{enumerate}
    \item  With band 7 single-pointing observation of GG Tau A, we achieved a spectral resolution of 0.1 km/s and an angular resolution of 0.8-0.9$''$. We spatially resolved the emission from \nnhp\ and \dcop\, while \hhdp, \CS, and \soo\ show no detection.
    \item Comparing the predictions of a parametric disk model to our observed visibilities along with the lower excitation transition observations obtained with NOEMA by \citet{Phuong2018} and \citet{Phuong2020}, we constrain the average temperature of \ce{N2H+} and \ce{DCO+} emitting regions to be around 12 K and 16 K respectively. This is at odds with the predictions from all the chemical models we explored and all the previous studies. Chemical models suggest \ce{DCO+} peaks at 20-25 K, where some \ce{CO} remains in the gas phase, and \ce{N2H+} peaks around 20 K, where \ce{CO} is fully frozen onto grains. On the contrary, our observational constraints indicate that both molecules trace regions below the CO layer \citep{Phuong2020a}, much closer to the midplane than previously thought, with \ce{N2H+} being the closest.
    \item Our radial distribution of \dcop\ emission reveals an atypical partially overlapping double ring structure. The inner ring peaks at $210 \pm 50$ au within the continuum ring region, while the more intense outer ring peaks at $375 \pm 10$ au in the outer disk. We hypothesize that \ce{DCO+} forms via two different pathways in the ring and the outer disk regions depending on gas-phase CO abundance. In the ring region, \ce{DCO+} forms through isotope exchange after \ce{HCO+} arises from \ce{CO} protonation. However, constraining \ce{DCO+} formation in the outer disk regions requires future work that takes into account the complex physical structure.

    \item \nnhp\ emission shows a ring-like radial distribution facilitated through protonation of \ce{N2} with its emission layer situated right below the \ce{DCO+} layer. The anti-correlation between \ce{N2H+} distribution with \ce{CO} distribution is clearly visible in our modelled simulation.
    
    \item We ran numerous chemical models varying CRI rates, C/O ratios, and stellar UV fluxes. These models suggest that the chemistry of \ce{H2D+}, \ce{N2H+}, and \ce{DCO+} in cold, dense regions is primarily influenced by gas-phase \ce{CO} abundance and the dominant midplane ion, \ce{H3+}. These regions are shielded from external UV radiation, making CRI the most crucial factor in determining \ce{CO} and \ce{H3+} abundances and, consequently, the distribution of \ce{N2H+}, \ce{DCO+}, and \ce{H2D+}. On the other hand, the chemistry of \ce{^{13}CS} and \ce{SO2} is very sensitive to the C/O ratio.
    
     \item After running a grid of astrochemical models to simultaneously fit the surface density constraints from the observed transitions along with the upper limits provided by the non-detections, our best-fit model, with a carbon-to-oxygen ratio of 1.0 and a sub-interstellar cosmic ray ionization rate of $10^{-18}$ s$^{-1}$, successfully reproduced the radial trends in surface density for \ce{DCO+} but slightly overestimated both the \ce{N2H+} content and the expected rotation temperatures. The CRI attenuation in the high-density GG Tau A ring can be attributed to an increased number of collisions experienced by CR particles as they gyrate towards the midplane.

    \item The non-detection of \hhdp\ suggests that ionization is much lower than the standard interstellar rate in the disk-midplane due to its very high density. In the cold midplane region, \ce{H2D+} is directly dependent on the abundance of \ce{H3+}, which, in turn, is regulated by the ionization rate. The non-detection of \ce{H2D+} aligns with our best model, where the expected surface density is below our observationally determined 3$\sigma$ upper limit only for a sub-interstellar CRI rate.
\end{enumerate}

The GG Tau A disk is very complex, but it is one of the nearest (hence larger) and more massive disks found around a low-mass star system. Such properties should facilitate the detection of \hhdp. However, its modelled surface density being close to the observationally constrained 3$\sigma$ upper limit, suggests that the detection of \ce{H2D+} is within the ALMA capabilities, although at the cost of a significantly longer integration time.

\section*{Acknowledgments}
This paper makes use of the following ALMA data: ADS/JAO ALMA$\#$2021.1.00342.S. ALMA is a partnership of ESO (representing its member states), NSF (USA) and NINS (Japan), together with NRC (Canada), MOST and ASIAA (Taiwan), and KASI (Republic of Korea), in cooperation with the Republic of Chile. The Joint ALMA Observatory is operated by ESO, AUI/NRAO and NAOJ. The National Radio Astronomy Observatory is a facility of the National Science Foundation operated under a cooperative agreement by Associated Universities, Inc. L.M. acknowledges the financial support from DAE and DST-SERB research grants (SRG/2021/002116 and MTR/2021/000864) of the Government of India. L.M. also acknowledges insightful discussions related to this project with Valentine Wakelam and Uma Gorti. A.D. and S.G. thank the French CNRS programs PNP, PNPS, and PCMI. This research was carried out in part at the Jet Propulsion Laboratory, which is operated for NASA by the California Institute of Technology. K.W. acknowledges the financial support from the NASA Emerging Worlds grant 18-EW-182-0083. Th.H. and D.S. acknowledge support from the European Research Council under the Horizon 2020 Framework Program via the ERC Advanced Grant Origins 83 24 28 (PI: Th. Henning). R.S. thanks NASA for financial support through various Astrophysics Data Program awards. N.T.P. acknowledges the financial support of Vingroup Innovation Foundation (VINIF) under project code VINIF.2023.DA.057$''$. A.C. received financial support from the European Research Council (ERC) under the European Union’s Horizon 2020 research and innovation programme (ERC Starting Grant “Chemtrip”, grant agreement No 949278). This work was also supported by the NKFIH excellence grant TKP2021-NKTA-64. We would like to thank the anonymous referees for their constructive comments, which helped improve the manuscript.

\facility{ALMA}   

\software{\texttt{CASA} \citep{casa}, \texttt{keplerian\_mask} \citep{kepmask}, \texttt{GoFish} \citep{GoFish}, \texttt{VISIBLE} \citep{loomis2018visible}, \texttt{dynesty} \citep{speagle2020dynesty}, \textsc{Imager} (\url{https://imager.oasu.u-bordeaux.fr}), \texttt{DNAUTILUS} \citep{Majumdar2017}, \texttt{numpy} \citep{numpy}, astropy \citep{astropy:2013, astropy:2018, astropy:2022}, \texttt{matplotlib} \citep{matplotlib}}


\appendix

\section{Channel Maps, Moment Maps and Radial Profiles}
\label{app:ch_maps}
\counterwithin{figure}{section}
\setcounter{figure}{0} 

\begin{figure}[ht]
    \centering
    \includegraphics[width=\textwidth]{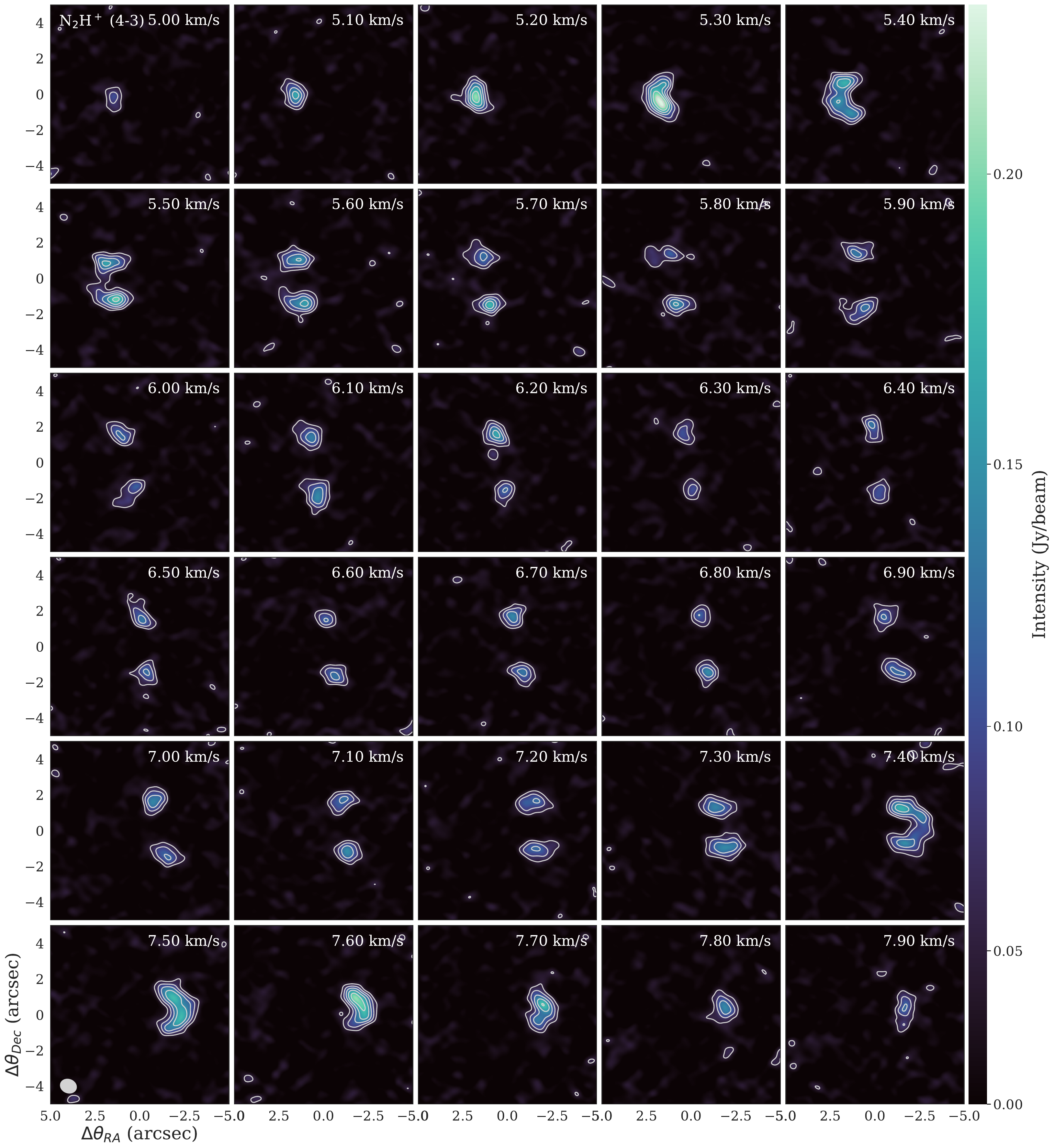}
    \caption{Channel maps of \nnhp\ imaged with \texttt{tclean} task in \texttt{CASA} as explained Section \ref{sec:obs}. The channel velocities (LSRK) are displayed in the top right corner of each channel in km/s. The solid contours are [3,5,7,9,...]$\times \sigma$ levels, where $\sigma = 14.6 $ mJy/beam. The ellipse at the lower left corner indicates the beam size of $0.93'' \times 0.77''$ with PA $68.76^\circ$. The colorbar is stretched non-linearly to showcase the line emissions more prominently.}
    \label{fig:ch_maps_n2hp}
\end{figure}

\begin{figure}
    \centering
    \includegraphics[width=\textwidth]{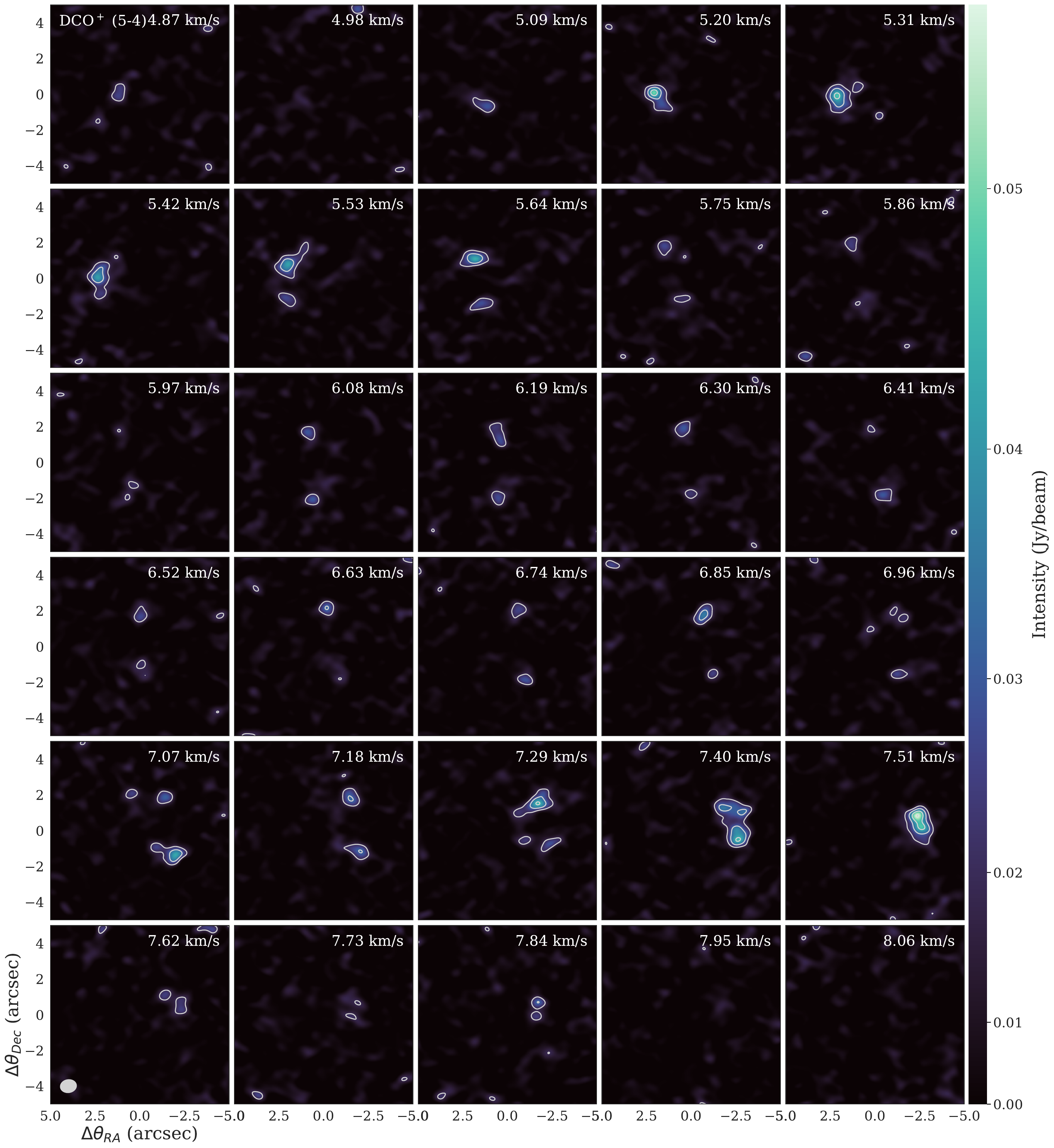}
    \caption{Channel maps of \dcop\ imaged with \texttt{tclean} task in \texttt{CASA} as explained Section \ref{sec:obs}. The channel velocities (LSRK) are displayed in the top right corner of each channel in km/s. The solid contours are [3,5,7,9,...]$\times \sigma$ levels, where $\sigma = 5.6 $ mJy/beam. The ellipse at the lower left corner indicates the beam size of $0.91'' \times 0.73''$ with PA $99.20^\circ$. The colorbar is stretched non-linearly to showcase the line emissions more prominently.}
    \label{fig:ch_maps_dcop}
\end{figure}

\begin{figure}
    \centering
    \includegraphics[width = \textwidth]{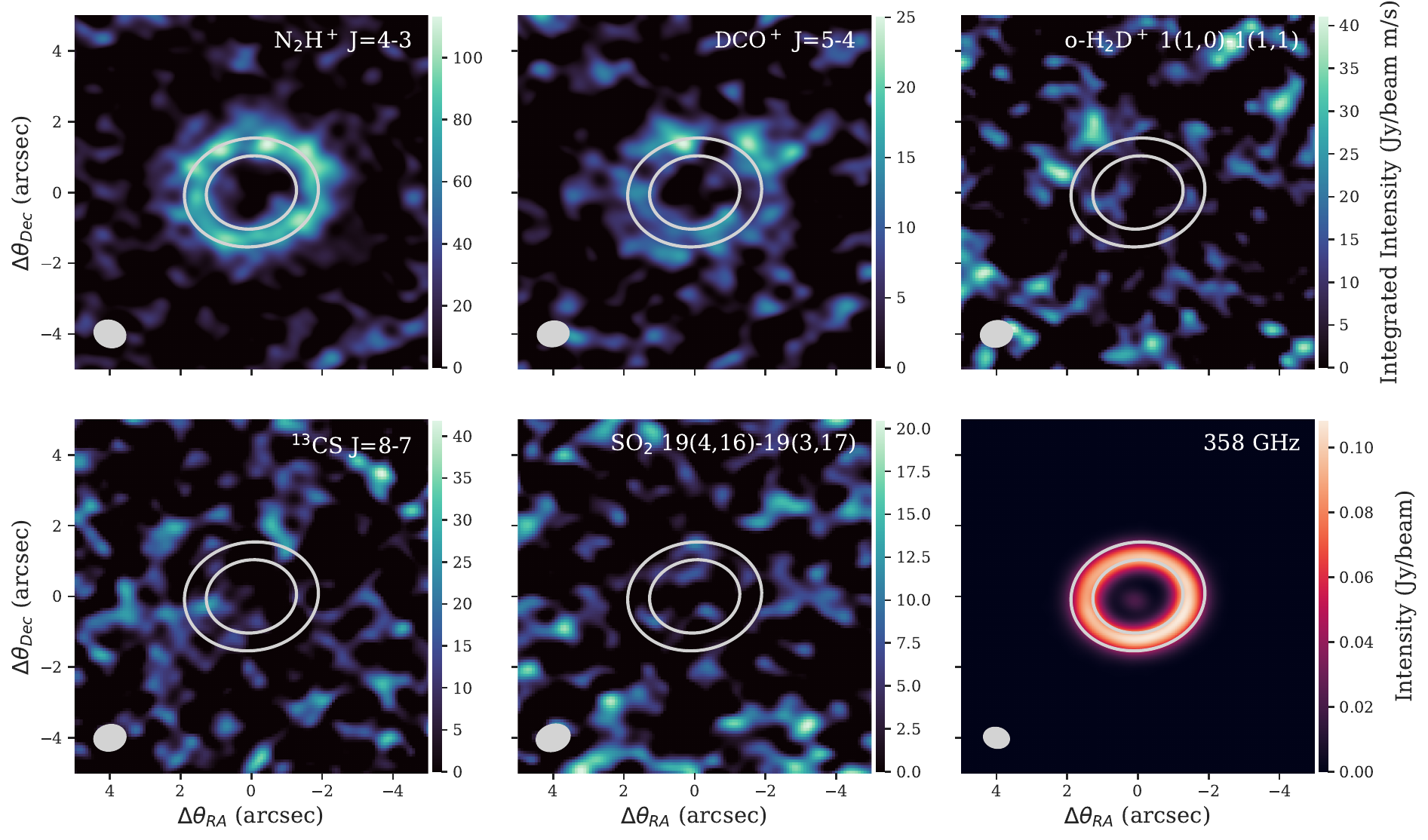}
    \caption{Integrated Intensity Maps of the targeted transitions generated without sigma clipping and masking. The overplotted elliptical ring contours in the white are at 193 au and 285 au to indicate the continuum ring region where 90\% of the circumtertiary emission arises \citep{Guilloteau1999}. The \textsg{panel at the bottom right} corner is the continuum image at 358 GHz.}
    \label{fig:m0_without_kepmask}
\end{figure}

\begin{figure}
    \centering
\includegraphics[width=\textwidth]{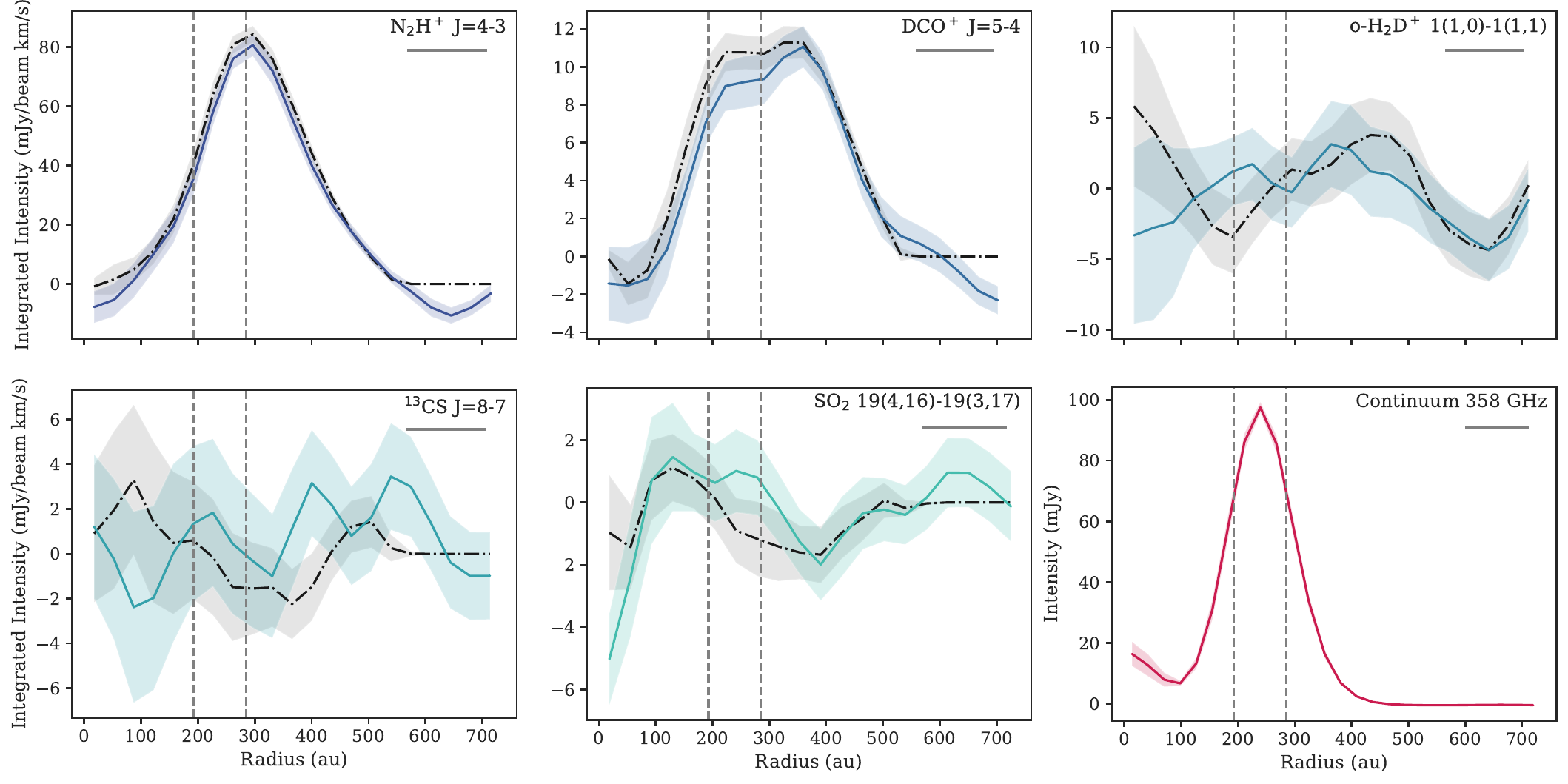}
    \caption{The dashed dot and solid curves represent azimuthally averaged velocity integrated radial profiles generated from integrated intensity maps with and without Keplerian masking, respectively. The shaded region in the above figure corresponds to 1$\sigma$ rms. The grey horizontal line at the top right denotes the beam major axis. The dashed vertical lines are to designate the location of the ring around GG Tau A, spanning from 193 au to 285 au. The subplot at the bottom right corner is the radial distribution of the azimuthally averaged continuum emission at 358 GHz.}
    \label{fig:rad_profile_wihtout_kep}
\end{figure}

\begin{figure}
    \centering
    \includegraphics[width=\textwidth]{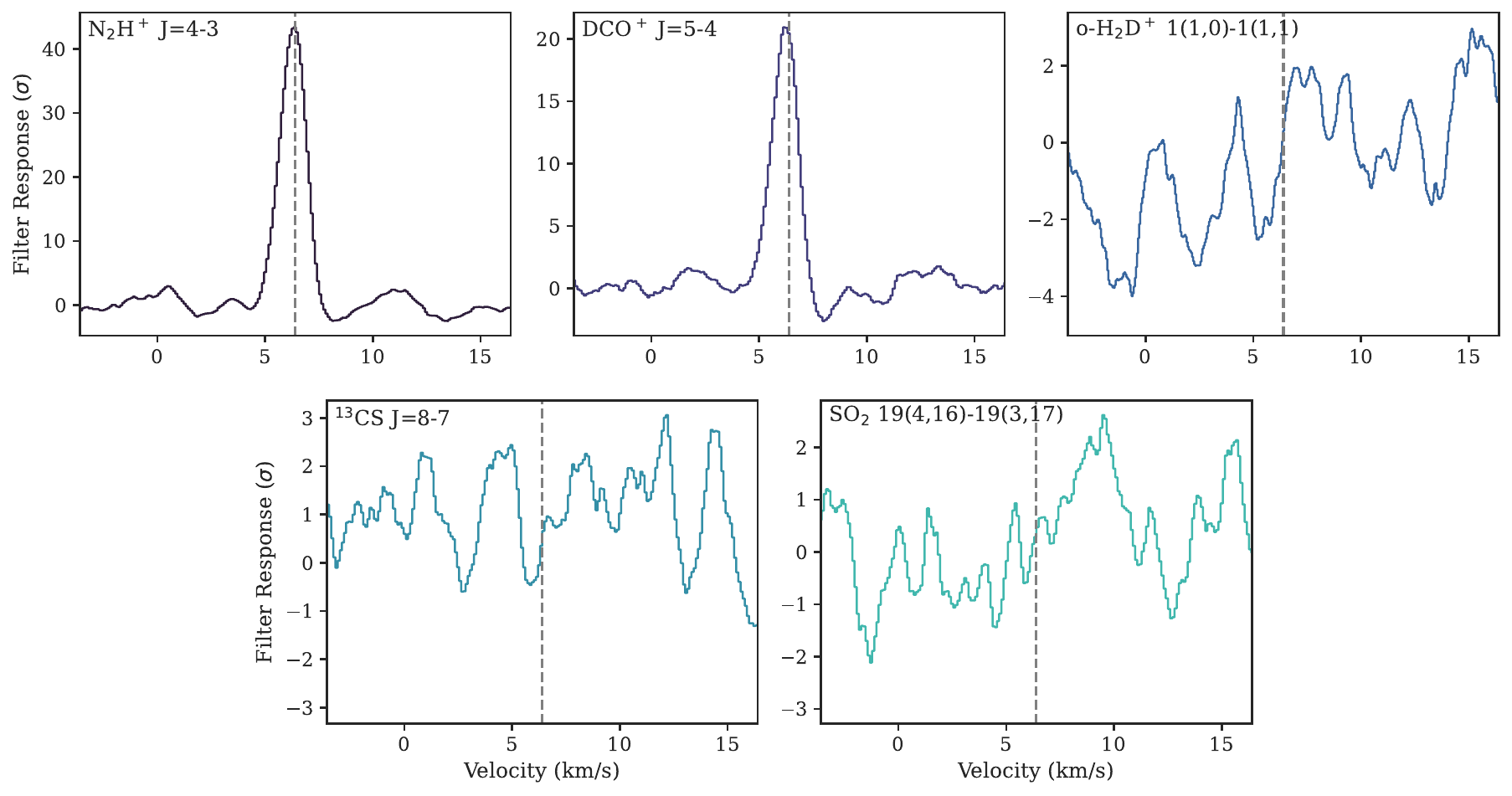}
    \caption{Matched filter responses to the emission for the targeted transitions, written at the upper left corner. The vertical grey dashed line is plotted at source velocity $v_{\text{LSRK}} = 6.4$ km/s for each subplot. Only \nnhp\ and \dcop\ show clear detection; rest of them do not.}
    \label{fig:matched_filter}
\end{figure}

The generated channel maps for the detected transition, \nnhp\ and \dcop\ are shown in Figures \ref{fig:ch_maps_n2hp} and \ref{fig:ch_maps_dcop}, respectively. Figure \ref{fig:m0_without_kepmask} shows integrated intensity maps between the velocity range 5.0 km/s and 8.0 km/s for all the targeted transitions. Sigma clipping and masking are not performed while generating these integrated intensity maps. Figure \ref{fig:rad_profile_wihtout_kep} shows the radial distribution of azimuthally averaged emission for the targeted transitions and the continuum generated from integrated intensity maps shown in Figure \ref{fig:m0_without_kepmask}. We have displayed the matched filter responses for all the targeted transitions in Figure \ref{fig:matched_filter}. The methods for generating all these figures are explained in Section \ref{subsec:obs_res}.

\section{IMAGER and DiskFit analyis} \label{app:imager}
\counterwithin{figure}{section}
\setcounter{figure}{0}

\begin{figure}    
    \centering
    \includegraphics[width=0.62\linewidth]{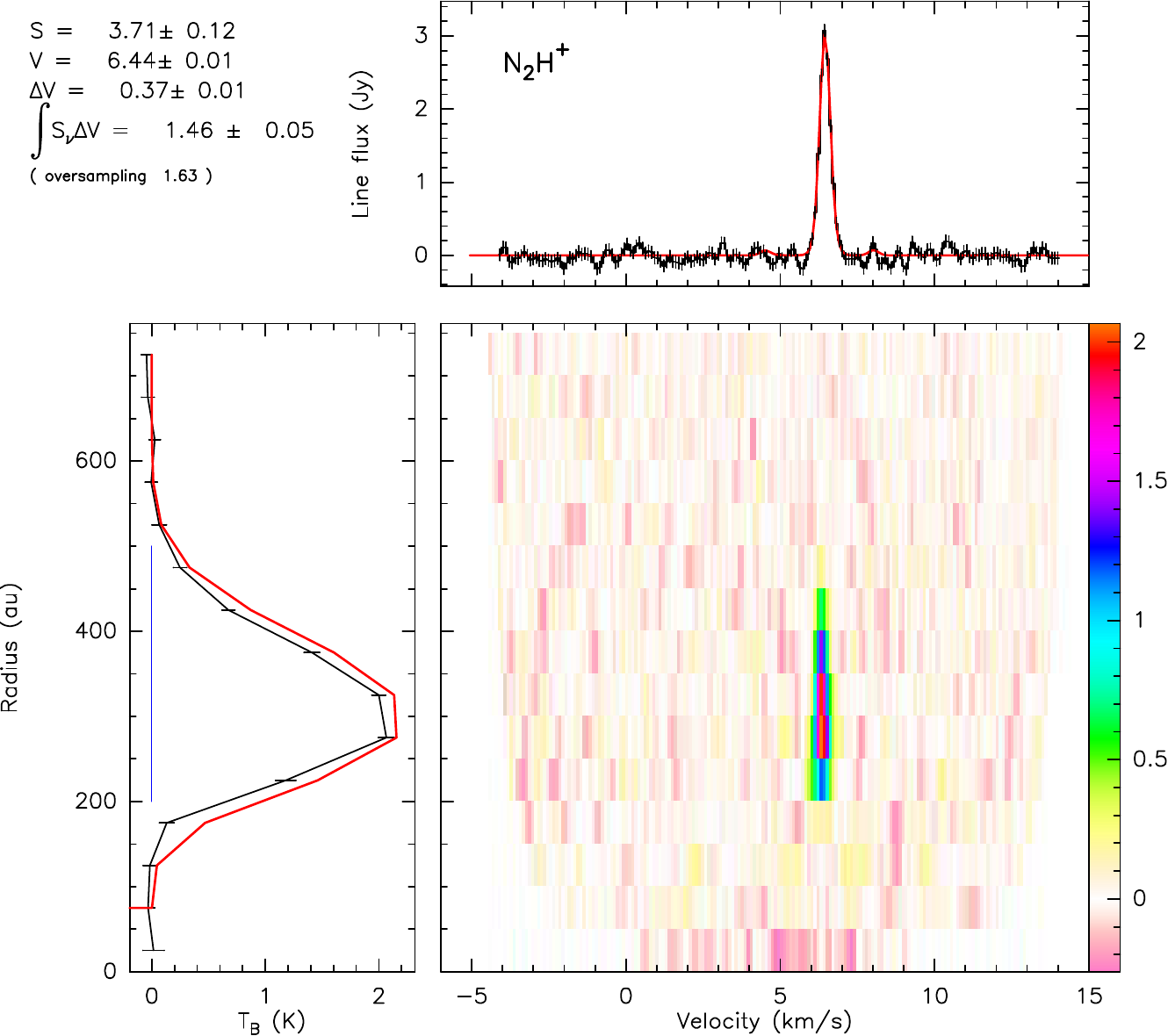}
    \\[\baselineskip]
    \includegraphics[width=0.62\linewidth]{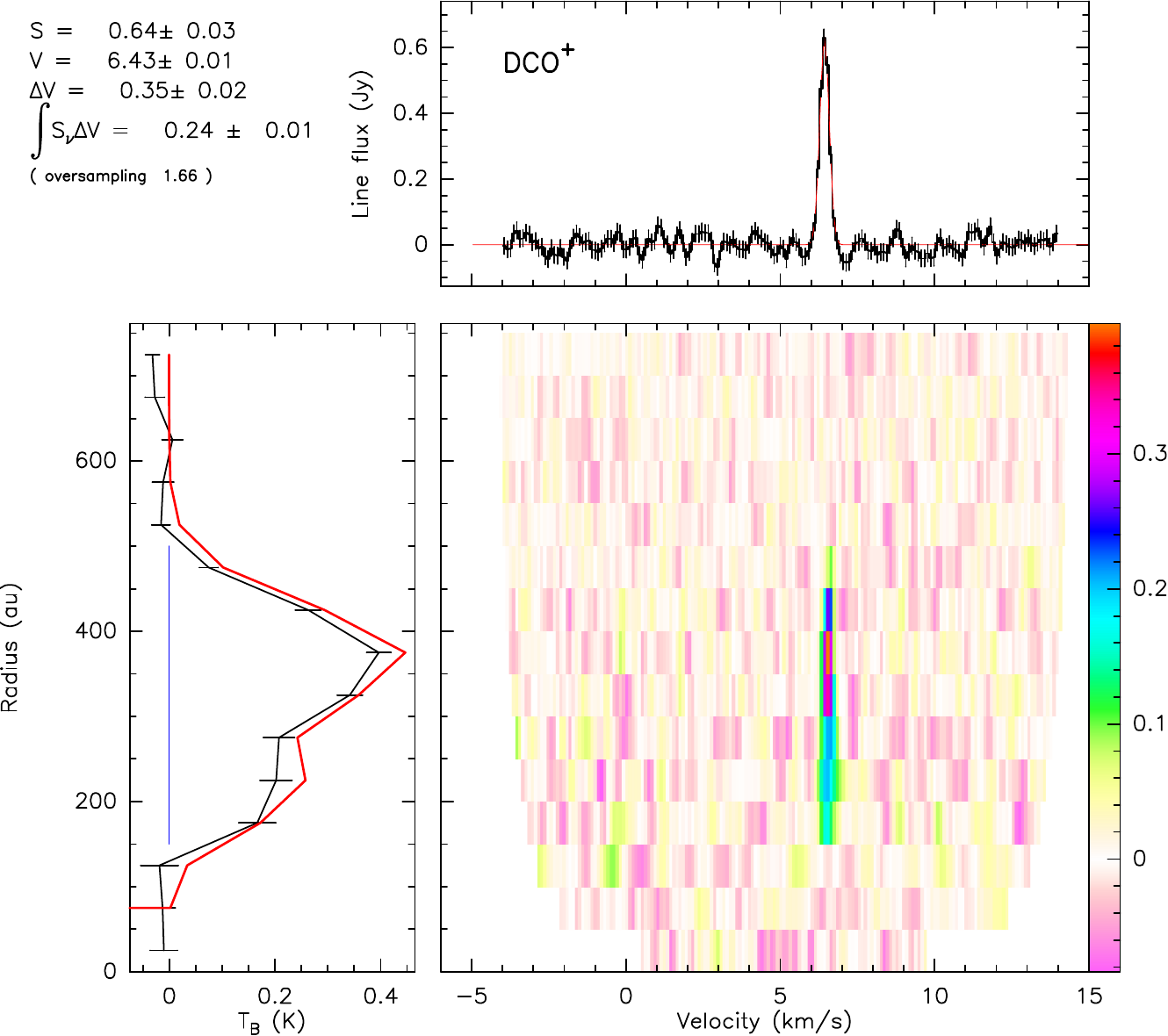}
\caption{Kepler plots generated for the observed transitions of \nnhp\ (top) and \dcop\ (bottom). For each molecule, the map is the brightness temperature (in K) as a function of velocity and radius
(``teardrop'' plot). The left panel shows the peak brightness radial profile, with error bars indicated, and the red profile is the one derived from the best-fit DiskFit model. The top panel shows the integrated spectrum over the range indicated by the blue line in the radial profile, with the best-fit Gaussian spectrum superimposed in red. The peak ($S$, in Jy) and integrated 
($\int S_\nu dV$, in Jy.km/s) flux densities, central velocity ($V$ in km/s), and FWHM line width ($\Delta V$ in km/s) obtained from a Gaussian fit to the integrated spectrum are given in the upper left corner of each panel.}

\label{fig:kepler-n2h}
\end{figure}

Figure \ref{fig:kepler-n2h} shows the
Keplerian deprojection produced by the \textsc{Imager} command \texttt{KEPLER}. Images used for this purpose were produced with \textsc{Imager}  and 
have an angular resolution of $0.66 \times 0.53''$
at PA $83^\circ$. Spectral resolution prior to
Keplerian velocity correction was about 0.1 km\,s$^{-1}$. This correction requires resampling of the spectra on a different grid, which results in correlated channels: the correlation factor is estimated from the \textsg{noise autocorrelation spectrum}.
The best fit radial profile derived from
\textsc{DiskFit} is always slightly above the measured one because the deconvolution with 
limited signal to noise cannot recover the full line brightness, while the $uv$ plane fitting method is not biased in this respect.

\section{Simple LTE Approach}
\label{sec:rotational_diagram}
We have undertaken a simpler approach to calculate disk-averaged surface densities where we assume local thermodynamic equilibrium (LTE) considering the disk densities are relatively high compared to critical densities for the studied transitions. This method requires prior knowledge of rotational temperature for a single transition.\par

For optically thin emission, the surface density of molecules in the upper state of a specific transition is expressed as,
\begin{align}
N_u^{thin} &= 
\frac{4\pi}{ h c A_{ul}} \frac{(\int S_\nu dv)}{\Omega} \\ 
&= 
\frac{8\pi k \nu^2 }{ h c^3 {A_{ul}}  } \left(\int T_b dv\right)
\end{align}
\noindent
where $\int S_\nu dv$ is the velocity integrated flux density, $A_{ul}$ is the Einstein coefficient, $\Omega$ denotes the solid angle subtended by the emission area, and \textit{h} and \textit{c} stand for the Planck constant and the speed of light in a vacuum, respectively \citep{goldsmith1999population}. In the alternate form,
$\nu$ is the line frequency and $T_b$ the line brightness temperature.

The total surface density can be derived from the upper state surface density, assuming a Boltzmann distribution
\begin{align}
    N_T^{thin} = N_u^{thin}  \frac{Q(T_{rot})}{g_u} exp \left(\frac{E_u}{T_{rot}}\right) \label{eq:Nt_thin}
\end{align}
\noindent
where $g_u$ is the degeneracy, $E_u$ is the energy of the upper energy level in K, and \textit{Q($T_{rot}$)} is the partition function at the rotation  temperature $T_{rot}$, that can be computed by summation
over the energy levels or by interpolating tabulated $Q$ values for discrete $T_{rot}$ values found in the Cologne Database for Molecular Spectroscopy (CDMS) \citep{endres2016cologne}.\par

If the transition is optically thick, the above value gives
a strict lower limit to the observed surface density. 
Opacity corrections are possible in a purely homogeneous medium with only turbulent or thermal line broadening \citep{goldsmith1999population}:
\begin{align}
    N_T = N_T^{thin} \frac{\tau}{1- e^{-\tau}} \label{eq:tau_correction}
\end{align}

where we can self-consistently calculate the optical depth, $\tau$ assuming a Gaussian-like line profile $\sim 1/\Delta v$ at the line center \citep[][]{goldsmith1999population}, $\Delta v$ = 0.25 km/s being the intrinsic full-width half maximum line width constrained from \textsc{DiskFit} analysis:
\begin{align}
\tau  = \frac{A_{ul}c^3}{8\pi \nu^3 \Delta v} N_u
    \left[ \exp\left(\frac{h \nu}{k T_{rot}} \right) - 1 \right]
    \label{eq:tau}
\end{align}

Unfortunately, in our case, the above-mentioned correction does not hold while calculating the radially resolved surface density, even if we apply this to radially dependent emission. Because of the Keplerian shear, there are significant regions in our images where the line of sight opacity varies both as a function of azimuth and radius, with both directions being insufficiently well resolved at our angular resolution (about 100 au). Accordingly, the azimuthally averaged brightness does not originate from a unique opacity below about 400 au.\par
However, the effect is expected to be small.
 The peak brightness of N$_2$H$^+$ (see Figure \ref{fig:kepler-n2h}) is 2\,K, while the brightness temperature of an optically thick line would be $J_\nu(T_{rot}) = 5.3$\,K for $T_{rot} = 12$\,K at the J=4-3 line frequency. This indicates a peak line opacity of about 0.47, leading to a $\sim 22$\% opacity correction at most. For DCO$^+$, whose emission is about 3 times fainter, the correction is even smaller, as for the (undetected) emission from H$_2$D$^+$ and other molecules. We thus expect the constraint given by the optically thin approximation to be a reasonable estimate for the mean surface density.\par

\begin{figure}
    \centering
    \includegraphics[width=\textwidth]{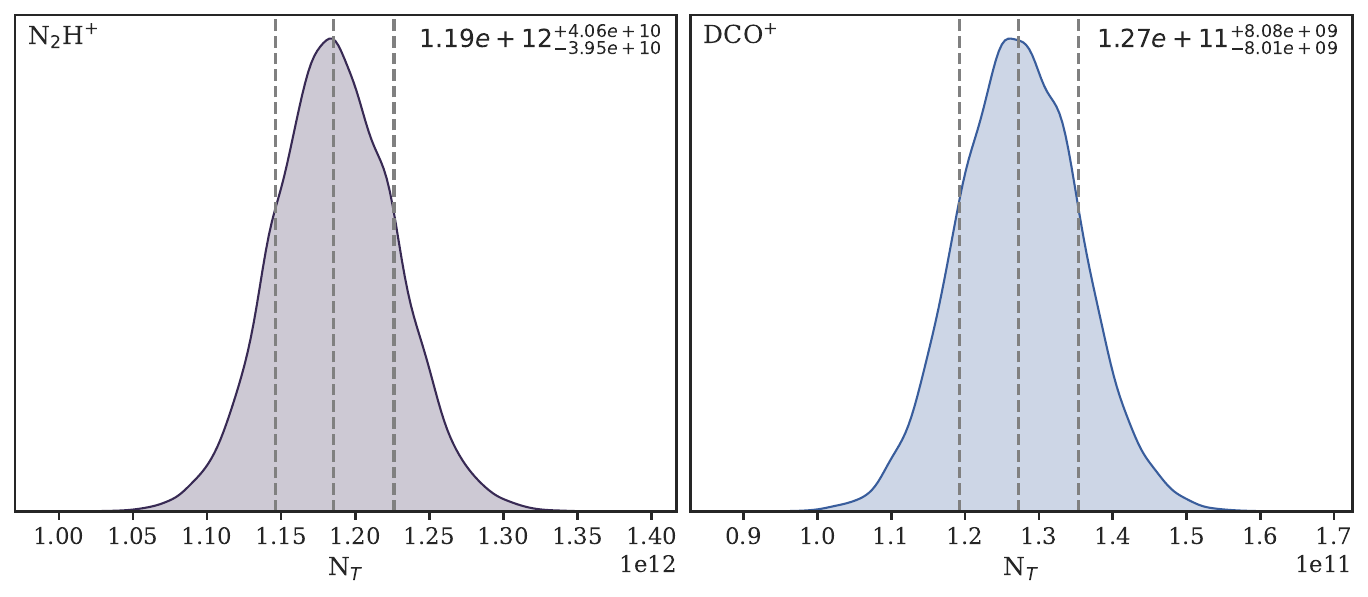}
    \caption{Posterior Distribution for disk averaged surface density calculations using the simple LTE approach. The left panel is for molecules \ce{N2H+}, and the right one is for \ce{DCO+}. The grey dashed lines represent 16, 50, and 84th percentile of the distribution. The median value with 1 $\sigma$ error bars (68\% confidence interval) are reported at the top right.}
    \label{fig:posterior}
\end{figure}

We explore the likely ranges of disk-averaged surface density using nested sampling while assuming log-normal priors for both the molecules in the range $10^5 <$ N$_{\text{T}} < 10^{18}$ cm$^{-2}$ and fixing the rotational temperature to the values constrained in \textsc{DiskFit} analysis. The rotation temperatures for N$_2$H$^+$ and DCO$^+$ were set at 12 K and 16 K, respectively. A value of 15 K was used for the undetected molecules. The Python package utilized to sample the posterior distribution is \texttt{dynesty} \citep{speagle2020dynesty}. We consider the median of the posterior as the best-fit value, while the 16th and 84th percentile values account for the $1 \sigma$ uncertainties. Posterior distributions for these averaged surface density derivations are shown in Figure \ref{fig:posterior}.

\section{Grids of Astrochemical Models and Finding the Best-Fit Model} \label{app:chemical_model}
\counterwithin{figure}{section}
\setcounter{figure}{0}

We ran a large number of astrochemical models with various combinations of parameters. The grids of models are summarized in Table \ref{tab:dnautilus_parameters}. Note that we have not implemented any grain growth in our models due to poor constraints on larger grain size. We explored the parameter space using the least square method while simultaneously considering the observational constraints on surface densities for \ce{N2H+} and \ce{DCO+} from our observations and $^{13}$CS as reported by Phuong et al. (2021). We ensured that the modelled values for undetected molecules remained below their respective upper limits. To quantify the goodness of fit, we calculated reduced chi-square values. This metric evaluates the agreement between observed and modelled surface densities across multiple molecules and radii. The reduced chi-square calculation is defined as follows:
\begin{align}
    \chi^2_{\text{red}} = \frac{1}{\nu} \sum_{j} \left( \sum_{i} \left( \frac{N_{\text{obs}}(i, j) - N_{\text{model}}(i, j)}{\sigma_{\text{obs}}(i, j) + \sigma_{\text{model}}(i, j)} \right)^2 \right)
\end{align}

In the equation, $\nu$, the number of degrees of freedom is 25. $N_{\text{obs}}(i, j)$ and $N_{\text{model}}(i, j)$ denote the observed and modelled surface densities of the j-th molecule at the i-th radial point, respectively. The terms $\sigma_{\text{obs}} (i,j) $ and $\sigma_{\text{model}} (i,j) $ represent the associated error estimates for the observed and modelled values, respectively. Observational errors were derived from the error bars in the peak column density values in Table \ref{tab:diskfitabun}, while modelled errors were assumed to be 20\% to account for uncertainties in the chemical models due to rate coefficients and other chemical parameters. The best-fit model is the one with the lowest $\chi^2_{\text{red}}$ value. Although the $\chi^2_{\text{red}}$ depends on the assumed errors, the best-fit solution was found to be independent of this assumption.


\begin{deluxetable*}{cccccc}[t]
\tablecaption{Parameter Space Explored with \text{DNAUTILUS}}
\label{tab:dnautilus_parameters}
\tablewidth{\textwidth}
\tablehead{
  \colhead{Initial Abundances} & \colhead{Settling} & \colhead{Grain Growth} & \colhead{$f_{\text{UV, 214 \text{ au}}}^a $ ($\chi_0$)} & \colhead{C/O$^b$} & \colhead{$\zeta_{\text{CR}}^c$ (s$^{-1}$)}
}
\startdata
Atomic$^d$ & no & no & 375, 1500, 3000 & 0.5, 0.7, 0.9, 1.0, 1.2 & $10^{-16}$, $10^{-17}$, $10^{-18}$, $10^{-19}$, $10^{-20}$ \\
$10^6$ yr Molecular Cloud$^e$ & no & no & 375, 1500, 3000 & 0.5, 0.7, 0.9, 1.0, 1.2 & $10^{-16}$, $10^{-17}$, $10^{-18}$, $10^{-19}$, $10^{-20}$ \\
Atomic$^d$ & yes & no & 375, 1500, 3000 & 0.5, 0.7, 0.9, 1.0, 1.2 & $10^{-16}$, $10^{-17}$, $10^{-18}$, $10^{-19}$, $10^{-20}$ \\
$10^6$ yr Molecular Cloud$^e$ & yes & no & 375, 1500, 3000 & 0.5, 0.7, 0.9, 1.0, 1.2 & $10^{-16}$, $10^{-17}$, $10^{-18}$, $10^{-19}$, $10^{-20}$ \\
\enddata
\tablecomments{
a) $f_{UV}$ is the stellar UV flux at reference radius, 214 au in the units of Draine interstellar UV field, $\chi_0$
b) Carbon to Oxygen (C/O) ratios have been considered around the standard value of 0.7 to account for ``low depletion" and ``high depletion" of oxygen \citep{Reboussin2015}.
c) Different cosmic ray ionization rates $\zeta_{\text{CR}}$ have been explored to account for how they impact our chemistry. d) In the reset scenario, we have taken our initial abundances in atomic forms along with \ce{H2} and \ce{HD} and run our astrochemical model for $10^6$ years. e) We have also considered the inheritance scenario as another possibility, where we have run a cloud model with atomic initial abundances, and the final abundances after $10^6$ years are used as initial abundance in the disk model. The disk model was then run for $10^6$ years.
}
\end{deluxetable*}

\clearpage
\bibliography{main}{}
\bibliographystyle{aasjournal}



\end{document}